\documentclass[prb,notitlepage,twocolumn,superscriptaddress,nofootinbib]{revtex4-2}
\usepackage{amssymb}
\usepackage{bbold}
\usepackage{hyperref}
\hypersetup{plainpages=false,colorlinks=true,linkcolor=blue, citecolor=blue, urlcolor=blue}
\usepackage{graphicx}
\usepackage{amsmath}
\usepackage{amssymb}
\usepackage{float}
\usepackage{mathtools}
\usepackage{multirow}
\usepackage{booktabs}
\usepackage{cleveref}
\usepackage{tikz}
\usepackage{physics}
\usetikzlibrary{shapes.geometric, arrows}
\pagecolor{white}
\tikzstyle{excitOp} = [rectangle, rounded corners, 
minimum width=3cm, 
minimum height=1cm,
text width = 3.5cm,
text centered, 
draw=black, 
fill=blue!25]
\tikzstyle{gateOp} = [rectangle, rounded corners, 
minimum width=2.5cm, 
minimum height=1cm,
text width = 3.5cm,
text centered, 
draw=black, 
fill=blue!25]
\tikzstyle{qcMeasure} = [rectangle, rounded corners, 
minimum height=1cm,
text width = 4cm,
text centered, 
draw=black, 
fill=red!25]
\tikzstyle{noiseFil} = [rectangle, rounded corners, 
minimum height=0.6cm,
text width = 2.3cm,
text centered, 
draw=black, 
fill=green!25]
\tikzstyle{greenFunc} = [rectangle, rounded corners,  
minimum height=1cm,
text width = 5.3cm,
text centered, 
draw=black, 
fill=green!25]
\tikzstyle{arrow} = [thick,->,>=stealth]
\newcommand{\ddotss}{%
\tikz[baseline]{
\filldraw (0ex,1.0ex) circle (.07ex);
\filldraw (.6ex,.5ex) circle (.07ex);
\filldraw (1.2ex,0ex) circle (.07ex);
}%
}
\renewcommand{\ket}[1]{|#1\rangle}
\renewcommand{\bra}[1]{\langle#1|}

\begin{document}

\title{Monte Carlo Diagonalization for Hubbard Model}
\author{B. Bernard}
\affiliation{D\'epartement de Biochimie, Chimie, Physique et Science Forensique, Institut de Recherche sur l'Hydrog\`ene, Universit\'e du Qu\'ebec \`a Trois-Rivi\`eres, Trois-Rivi\`eres, Qu\'ebec G9A 5H7, Canada}
\author{M. Charlebois}
\affiliation{D\'epartement de Biochimie, Chimie, Physique et Science Forensique, Institut de Recherche sur l'Hydrog\`ene, Universit\'e du Qu\'ebec \`a Trois-Rivi\`eres, Trois-Rivi\`eres, Qu\'ebec G9A 5H7, Canada}

\begin{abstract}
The Hubbard model has often been studied with exact diagonalization (ED). This impurity solver is fundamentally limited by the exponential scaling of the Fock space.
To address this problem, we introduce Monte Carlo diagonalization. Using a truncated Fock space constructed with a Monte Carlo approach, we reduce the size of the basis required to represent the Hamiltonian. We can then apply the Lanczos and band Lanczos algorithms in this truncated basis to find the ground state and the Green function.
This results in a significant economy of resources, an important speed up in the computational time and the capacity to break the $\sim 20$-site limit of ED. 
\end{abstract}
\maketitle
\section{Introduction}

The Hubbard model~\cite{hubbard-hamiltonian} is the simplest model to capture the behavior and properties emerging from strongly correlated electron materials.
The competition between the kinetic energy and the Coulomb repulsion results in a variety of interesting phases. It is often used to study cuprates because, combined with the right numerical method, it is possible to observe the different phases seen in cuprates (Mott, pseudogap, superconductivity, antiferromagnetism, and Fermi liquid).

There are many numerical methods developed to attempt to solve strongly correlated electron models~\cite{Qin_Schafer_Andergassen_Corboz_Gull_2022, LeBlanc2015,Schaefer2021}. Many quantum cluster methods and quantum impurity solvers have been developed with different approaches. Exact Diagonalization (ED)~\cite{Caffarel:1994,dagotto} is exact but limited to finite baths, continuous-time quantum Monte Carlo (CTQMC)~\cite{RMP_CTQMC,Haule2007} works at finite temperature and allows for infinite baths but is limited by the sign problem and requires ill-defined analytical continuation. Variational Monte Carlo (VMC)~\cite{PhysRev.138.A442,PhysRevB.71.241103,VMC2008,Charlebois2020,Rosenberg2022} has polynomial scaling (instead of exponential) but is limited to a biased ansatz. There is also the density matrix renormalization group (DMRG)~\cite{PhysRevLett.69.2863} and its extensions, diagrammatic Monte Carlo (DiagMC)~\cite{VANHOUCKE201095}, to name a few.

In order to reach the thermodynamic limit to study spontaneous symmetry breaking and to explore the phase diagram of cuprates, it is often necessary to embed these impurity solvers to reproduce infinite lattices. Many embedding techniques exist, such as Cluster Perturbation Theory (CPT)~\cite{gros_cluster_1993, senechal_spectral_2000,Senechal2002}, Cluster Dynamical Mean-Field Theory (CDMFT)~\cite{cdmft-lichtenstein,cdmft-kotliar}, Dynamical Cluster Approximation (DCA)~\cite{maier_quantum_2005-3}, TRILEX~\cite{PhysRevB.92.115109}, and D$\Gamma$A~\cite{PhysRevB.75.045118}, for example. 

VMC is particularly impressive because it can capture the physics of Fermi arcs on 64-site clusters using only a few million samples of the Fock space~\cite{Rosenberg2022}. What if, instead, we do not use an ansatz and only keep these states in order to diagonalize the resulting Hamiltonian matrix? In this article, we explore this hypothesis. We introduce \textit{Monte Carlo diagonalization}, which is a simple optimization of the ED solver. This method consists of building a basis of the most important Fock states, using Monte Carlo sampling, and then diagonalizing the resulting Hamiltonian matrix in this truncated basis.

Basis truncation has been extensively explored in quantum chemistry to study electron interactions in different shells of an atom. Configuration interaction (CI) uses linear expansions of the wave function in terms of Slater determinants generated by applying excitations and keep the most important states. CI has seen a revival recently through many different truncation schemes~\cite{sherrill1999,tubman2016-2,holmes2016,schriber2016,schriber2017,zimmerman2017,zimmerman2017-2,tubman2020}. Truncated CI has also been combined with dynamical mean-field theory (DMFT) to study the Hubbard model~\cite{Zgid_2012,Go_2017,mejuto2019}. 

A similar method is Quantum Monte Carlo diagonalization (QMCD)~\cite{otsuka1999,takashi2007}, which constructs a truncated basis of Slater determinants corresponding to the configuration of auxiliary fields. QMCD is very similar to \textit{Quantum Monte Carlo} (QMC)~\cite{PhysRevLett.51.1900,PhysRevB.31.4403,takashi2007} where the groundstate is $\ket\Omega = e^{-\tau H} \ket{\psi_0}$ and where $\ket{\psi_0}$ is the non-interacting Fermi sea. $\ket\Omega$ becomes exact for $\tau\rightarrow\infty$. This calculation can be performed using a Suzuki-Trotter decomposition followed by Hubbard-Stratonovich transformation, resulting in a decomposition over the different Slater determinants of the configurations of the auxiliary fields $\{\ket{\phi_n}\}$. This basis is too large for the calculation to be performed; thus it needs to be sampled with Monte Carlo. In QMCD, the truncated basis is used to express the Hamiltonian in the non-orthogonal subspace defined by $\{\ket{\phi_n}\}$, and can be diagonalized without any sign problem.


QMCD is similar in spirit to MCD: the method introduced in this article, hence the similarity in their names. However, the basis sampled is completely different. Compared to most of the methods cited above, MCD is relatively simple to implement. It requires only minor modifications to ED and makes it possible to reach cluster sizes beyond the $\sim 20$-site limit of ED~\cite{toyama2005}. This speedup opens the door to using the ED impurity solver within CDMFT for larger clusters than those accessible with ED alone.

The article is organized as follows. We first present the Hubbard model in Sec.~\ref{sec:Model}. For completeness, we summarize the ED algorithm in Sec.~\ref{sec:ED} to motivate our method. We introduce Monte Carlo diagonalization in Sec.~\ref{sec:Method}. We show benchmarks and results in Sec.~\ref{sec:Results} and discuss them in Sec.~\ref{sec:discussion}. An open-source version of the code developed for this work can be found in the supplemental material.

\section{Model}\label{sec:Model}
The Hubbard model is given by:
\begin{equation}\label{eq:hubbard_model}
 \hat{H} = -t \sum_{ij\sigma} \hat{c}_{i\sigma}^\dag \hat{c}_{j\sigma} + U \sum_{i} \hat n_{i\uparrow} \hat n_{i\downarrow} - \mu \sum_{i\sigma} \hat n_{i\sigma},
\end{equation}
where $\hat c_{i\sigma}^\dag$ ($\hat c_{i\sigma}$) creates (annihilates) an electron of spin $\sigma$ on site $i$. $\hat n_{i\sigma} = \hat c_{i\sigma}^\dag \hat c_{i\sigma}$ is the number operator. $t$ is the nearest-neighbor hopping amplitude, $U$ is the on-site Coulomb repulsion, and $\mu$ is the chemical potential. 
%
\section{Exact diagonalization}\label{sec:ED}

\begin{figure}[t]
    \includegraphics[width=\columnwidth]{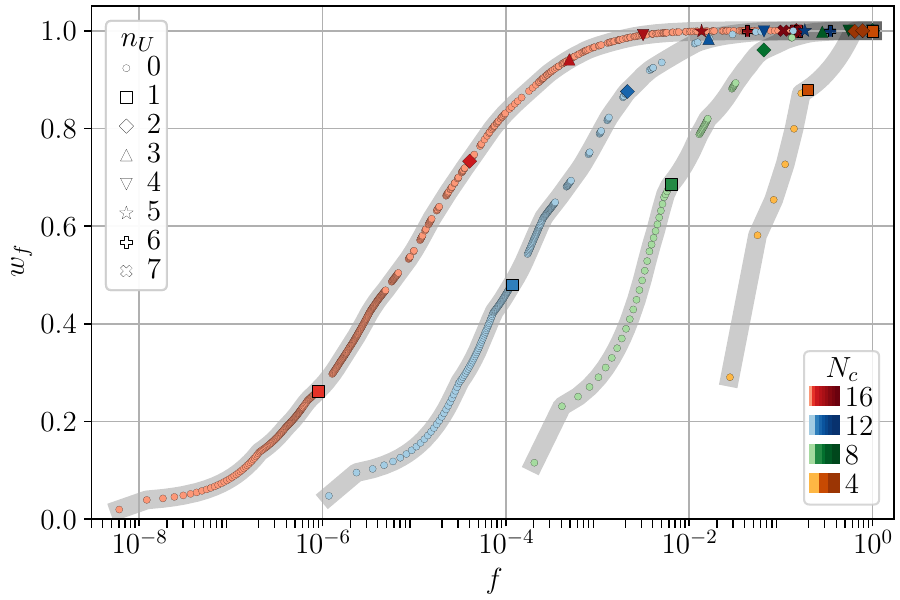}
    \caption{The cumulative combined weight $w_f$ of the $N_f$ most important Fock states as a function of $f$ in 1D, using 4-, 8-, 12-, and 16-site linear clusters at half-filling. Here $U=8$, $t=1$, $\mu=4$, and the sector is $N_e=N_c$ and $S_z=0$. The shape of the symbols is associated with the double occupation of a state $n_U$. To improve readability, we only render the state with the highest weight per $n_U$ when $n_U>0$. Even for $n_U=0$, states too close to each other were omitted. We used periodic boundary conditions in 1D.}
    \label{fig:energy_conv_1D}
\end{figure}

The Hubbard model can be solved using exact diagonalization~\cite{dagotto,Caffarel:1994} on small clusters of size $N_c$ sites. We can express the Hamiltonian $\hat{H}$ as a matrix $\mathbf{H}$ using the Fock space $\{\ket{x_n}\}$:
\begin{align}\label{eq_:H_matrix_full_space}
    H_{mn} = \mel{x_m}{\hat{H}}{x_n}.
\end{align}
The Fock space has $4^{N_c}$ states and thus grows exponentially, which is a major hurdle for this method.

Since the Hubbard model conserves the number of electrons $N_e$ and the $z$-projection of the spin $S_z$:
\begin{align}
[\hat H,\hat{N}_e]&=0
&
[\hat H,\hat{S}_z]&=0
\end{align}
the resulting matrix $\mathbf{H}$ is block diagonal. A block with $N_e$ electrons and spin $S_z$ has a total of $N_{\text{tot}}=C^{N_c}_{N_{e\uparrow}} C^{N_c}_{N_{e\downarrow}}$ states, where $C$ is the combination symbol. We can choose any of these blocks and diagonalize the matrix using the Lanczos algorithm to efficiently calculate the ground state energy $\Omega$ and the ground state vector $\ket\Omega$. The Lanczos algorithm~\cite{lanczos_iteration_1950} is described in Appendix~\ref{app:lanczos_algorithm}. In the end, the ground state $\ket\Omega$ is expressed as a collection of coefficients of each Fock state $\ket{x_i}$:
\begin{align}
\ket\Omega = \sum_n c_n \ket{x_n}.
\end{align}
The contribution of a specific Fock state $\ket{x_n}$ to the ground state is given by $\vert c_n \vert^2$. Suppose now that we are interested in the partial weight of only a number of states $N_f$. Then this partial weight is: 
\begin{align}
w_f = \sum_n^{N_f} \vert c_n \vert^2.
\end{align}
We also define $f =
N_f/N_{\text{tot}}$, i.e., the fraction of states corresponding to $N_f$.

\begin{figure}[t]
    \includegraphics[width=\columnwidth]{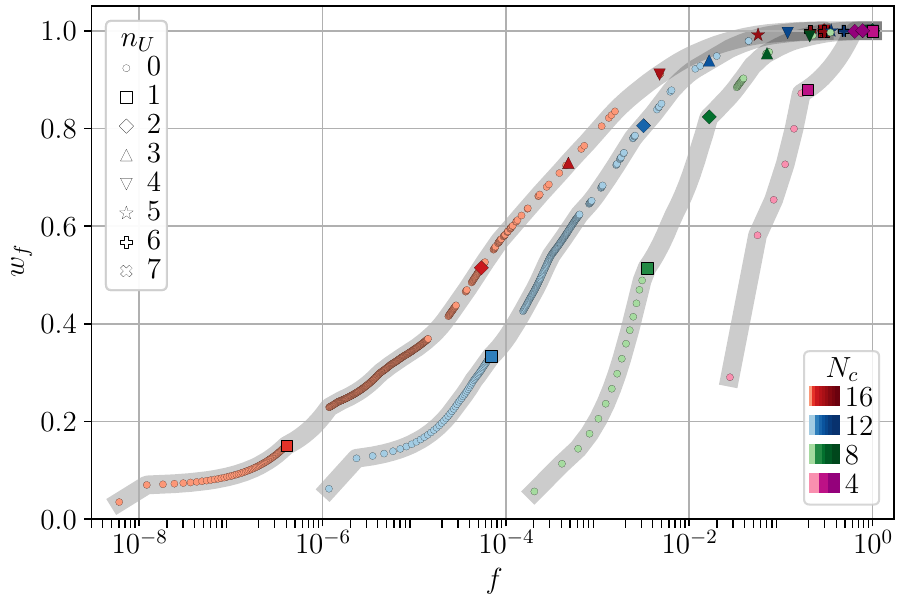}
    \caption{Same as Fig.~\ref{fig:energy_conv_1D}, but in a 2D lattice. Here, $N_c=4$ is a $2\times 2$ cluster, $N_c=8$ is $2\times 4$, $N_c=12$ is $3\times 4$, and $N_c=16$ is $4\times 4$. We used periodic boundary conditions in 2D.}
    \label{fig:energy_conv_2D}
\end{figure}

After $\ket\Omega$ is obtained from the Lanczos algorithm, the $c_n$ can be sorted from highest to lowest to study the weight $w_f$ of the most important Fock states. The $w_f$ corresponding to the most important Fock states as a function of $f$ for many clusters on the 1D and 2D Hubbard models at half-filling, in the block $N_e=N_c$ and $S_z=0$, is presented in Figs.~\ref{fig:energy_conv_1D} and~\ref{fig:energy_conv_2D}, respectively. 

We see that $w_f$ increases as $f$ increases, as expected, but also that $w_f \sim 1$ for $f \ll 1$. These results show that the fraction of states required to reach a combined weight of $0.99$ is very low, and the larger the cluster becomes, the smaller this fraction becomes. 

The double occupation number of $\ket{x_n}$, defined as: 
\begin{equation}
    n_U=\bra{x_n}\sum_i \hat{n}_{i\uparrow} \hat{n}_{i\downarrow}\ket{x_n}
\end{equation}
is also shown in Figs.~\ref{fig:energy_conv_1D} and~\ref{fig:energy_conv_2D}. The $n_U=0$ states are drawn, but only the first and most important state where $n_U > 0$ is presented, to enable readability. Clearly, the most important states have the least double occupation. However, it is interesting that some $n_U=0$ states are much less important than $n_U=1,2,3$, and even $n_U=4$ in 1D. This effect is more pronounced in 1D than in 2D. This is because some states, where most electrons are bunched together, are much less probable in 1D and thus contribute much less to the ground state.

The key idea is that only a fraction of states $f$ is important to capture all the physics of the ground state. If it were possible to know which states are more important, this could reduce the necessity for the full Fock space, speed up calculations, and allow for larger clusters. Of course, we do not know how ``important'' a state is before knowing the ground state. However, we can follow the intuition given by Figs.~\ref{fig:energy_conv_1D} and~\ref{fig:energy_conv_2D} and prioritize low $n_U$ states, as a first approximation.

\section{Monte Carlo Diagonalization}\label{sec:Method}
In this section, we introduce a way to construct two suitable subspaces of Fock space: one for the calculation of the ground state and the other for the calculation of the Green function.

\subsection{Subspace of the ground state}\label{sec:sampling}

Let $\mathcal{F}$ be the set containing every state of the Fock space within a sector with $N_e$ and $S_z$. We define the subspace $\tilde{\mathcal{F}}$ as a subset of $\mathcal{F}$:
\begin{equation}
\begin{aligned}
    {\mathcal{F}} &=\{\underbrace{\ket{x_1},\ket{x_2},...,\ket{x_{N_f}}},...,\ket{x_N}\}\\
    \tilde{\mathcal{F}} &= \{\ket{\tilde{x}_1},\ket{\tilde{x}_2},...,\ket{\tilde{x}_{N_f}}\}
\end{aligned}
\end{equation}
Here $\ket{x_n}$ and $\ket{\tilde{x}_n}$ are the same state, but the \textit{tilde} symbol is used to make it clear that it belongs to the subset of ``important states'' $\tilde{\mathcal{F}}$. In fact, all the quantities noted with a \textit{tilde} in the article refer to those associated with the truncated basis $\tilde{\mathcal{F}}$.

The Hamiltonian matrix $\mathbf{\tilde H}$ of this subspace $\tilde{\mathcal{F}}$ is obtained by: 
\begin{equation}\label{eq:h_matrix_projected}
 \tilde H_{mn} = \mel{\tilde{x}_m}{\hat H}{\tilde{x}_n},
\end{equation}
and its ground state energy $\tilde{\Omega}$ and the ground state vector $\ket{\tilde\Omega}$ obtained using the same Lanczos algorithm (see Appendix~\ref{app:lanczos_algorithm}).

Our proposal to construct $\tilde{\mathcal{F}}$ is to start with one very probable state, such as one of the two antiferromagnetic (AFM) states at half-filling, and to generate new states by applying every hopping included in the kinetic part of the Hubbard model:
\begin{equation}\label{eq:hopping_operator}
\sum_{\langle ij\rangle\sigma} \hat{c}_{i\sigma}^\dag\hat{c}_{j\sigma}.
\end{equation}
The suggested states are accepted in $\tilde{\mathcal{F}}$ with probability:
\begin{equation}\label{eq:boltzmann_factor}
P_{a \rightarrow b} = \min\Big(\exp\big({-U\beta {n_U}_b}\big),1\Big),
\end{equation}
where ${n_U}_b$ is the double occupation of the suggested state labeled $b$, and $\beta = 1/k_BT$. 

We observed that it is optimal to favor states that are close to each other. 
To do so, 
it is better to propose new states by \textit{steps} as follows:
\begin{enumerate}
  \item In one \textit{step}, only the newly added states of the last \textit{step} act as starting points to search for new important states. 
  \item We apply every possible hopping to each newly added state of the last \textit{step}. This generates many proposed states.
  \item The proposed states are accepted with probability $P_{a \rightarrow b}$ of Eq.~\eqref{eq:boltzmann_factor}. Every accepted state in this \textit{step} is labeled as ``newly added'' and will act as a starting point for the next \textit{step}.
  \item This is repeated until all newly added states of the last \textit{step} are processed. We then proceed to the next \textit{step}.
\end{enumerate}
%
We proceed this way until the desired number of states $N_f$ (or $f$) is reached.
It is important that states that have already been accepted and visited can be proposed in future iterations. These redundant propositions do not contribute to the growth of $\tilde{\mathcal{F}}$, but they allow us to revisit very probable states and populate any missing states. This ensures the ergodicity of the Monte Carlo process. Without this possibility, $\tilde{\mathcal{F}}$ would contain groups of states impossible to reach after some sampling time. 
This approach, where each newly added state is tested with every hopping term, results in a search more akin to breadth-first search and yields better results than a more random-walk or depth-first search approach. See the discussion for interpretation.

The search can be long and cumbersome if we are not careful, for large systems. It is useful to encode $\tilde{\mathcal{F}}$ in a binary tree to store and search the most important states already included.
The search is especially long for a large fraction of states (above $f=0.5$ for example) in large clusters. We can speed up this process greatly by pre-accepting all the states with lower $n_U$, and performing the Monte Carlo search in the remaining $n_U$ only. Finally, it is important that $\beta$ is large enough to favor low double occupation states, but small enough so that the proposed states with higher $n_U$ are not always rejected.

\subsection{Subspace of the Green function} \label{sec:GreenFunction}
The Green function can be expressed in the Lehmann representation as:
\begin{align}
  G_{\mu\nu}(\omega) &= G^{+}_{\mu\nu}(\omega) + G^{-}_{\mu\nu}(\omega)\label{eq:green}\\
  G^{+}_{\mu\nu}(\omega) &= \bra{\tilde\Omega}\hat{c}_{\mu} \frac{1}{\omega +i\eta +  \tilde{\Omega} - \hat H} \hat{c}_{\nu}^\dag\ket{\tilde\Omega}\label{eq:green_plus}\\
  G^{-}_{\mu\nu}(\omega) &= \bra{\tilde\Omega}\hat{c}_{\mu}^\dag \frac{1}{\omega +i\eta - \tilde{\Omega}  + \hat H} \hat{c}_{\nu}\ket{\tilde\Omega},
  \label{eq:green_minus}
\end{align}
where the index $\mu$ is shorthand for both the site index $i$ and the spin index $\sigma$:
\begin{equation*}
    \mu :=\{i,\sigma\}.
\end{equation*}
$\eta$ is the Lorentzian broadening.
As we did for the ground state calculation, we also want to reduce the fraction of states necessary to compute the Green function, and thus we need to construct a truncated basis of the excited sectors. 
We construct $\tilde{\mathcal{F}}^\pm$ by applying every $\hat{c}_{\mu}^{(\dag)}$ to every state in $\tilde{\mathcal{F}}$ and keeping each state only once:
\begin{equation}
\begin{aligned}
\tilde{\mathcal{F}}^+ &= \bigcup_\mu \hat{c}_{\mu}^\dag\tilde{\mathcal{F}}  = \bigcup_\mu \qty{\hat{c}_{\mu}^{\dag}\ket{\tilde{x}_1},...,\hat{c}_{\mu}^{\dag}\ket{\tilde{x}_{N_f}}} 
\equiv
\qty{\ket{\tilde{x}_n^+}} 
\\
\tilde{\mathcal{F}}^- &= \bigcup_\mu \hat{c}_{\mu}\tilde{\mathcal{F}} = \bigcup_\mu \qty{\hat{c}_{\mu}^{\phantom{\dag}}\ket{\tilde{x}_1},...,\hat{c}_{\mu}\ket{\tilde{x}_{N_f}}}
\equiv
\qty{\ket{\tilde{x}_n^-}}.
\end{aligned}
\label{eq:excited_subspace}
\end{equation}
Using these subspaces, we can find the Green function with the band Lanczos algorithm. Details of the band Lanczos algorithm can be found in Appendix~\ref{app:band_lanczos_algorithm}. We use band Lanczos to 
approximate the matrices of the excited sectors $\mathbf{\tilde{H}^\pm} = \mel{\tilde x_m^\pm}{\hat H}{\tilde x_n^\pm}$ as effective matrices of lower order $\tilde{\mathbf{H}}_{\rm{eff}}^\pm$ and we diagonalize these effective matrices. We then use the eigenpairs ($\tilde{E}_n^\pm$,$|\tilde{E}_n^\pm\rangle$) of  $\tilde{\mathbf{H}}_{\rm{eff}}^\pm$ to rewrite the Green function as:
\begin{align}\label{eq:green_qmatrix}
 G^{+}_{\mu\nu}(\omega) &=\sum_m\frac{\tilde{Q}^+_{\mu m}\tilde{Q}^{+*}_{\nu m}}{\omega +i\eta + \tilde\Omega - \tilde{E}^+_m},
\\
 G^{-}_{\mu\nu}(\omega) &=\sum_n\frac{\tilde{Q}^-_{\mu n}\tilde{Q}^{-*}_{\nu n}} {\omega +i\eta - \tilde\Omega + \tilde{E}^-_n},
 \label{eq:green_qmatrix_m}
\end{align}
where
\begin{align}\label{eq:q_matrix}
 \tilde{Q}^+_{\mu m} = \langle\tilde{\Omega}|\hat{c}_{\mu}|\tilde{E}^+_m\rangle, && 
 \tilde{Q}^-_{\mu n} = \langle\tilde{\Omega}|\hat{c}_{\mu}^\dag|\tilde{E}^-_n\rangle.
\end{align}
If the ground state energy is degenerate, we repeat the same $G_{\mu\nu}(\omega)$ calculation for every ground state, and sum them together. 

In Sec.~\ref{sec:Results}, particular emphasis is given to the total density of states (DOS) calculated from the Green function:
\begin{equation}\label{eq:DOS}
 A(\omega) = -\frac1\pi \Im \Big[ \sum_\mu G_{\mu\mu}(\omega)\Big].
\end{equation}

\subsection{Further truncation}\label{sec:truncation}
In Eq.~\eqref{eq:excited_subspace}, we see that there is an increase in size when going from the subspace of the ground state $\tilde{\mathcal{F}}$ to the excited subspaces $\tilde{\mathcal{F}}^\pm$. Without further truncation, the size of these excited subspaces $\tilde{\mathcal{F}}^\pm$ would be the computational bottleneck. 

Fortunately, there is a simple remedy, because after we find $\ket{\tilde\Omega}$, we know the most important states of $\tilde{\mathcal{F}}$, since we know the coefficients of $\ket{\tilde\Omega}$. As a last refinement, we can further truncate the subspace $\tilde{\mathcal{F}}$ after the calculation of the ground state $\ket{\tilde\Omega}$ and before calculating $\tilde{\mathcal{F}}^\pm$. We simply normalize $\ket{\tilde\Omega}$, order the coefficients of $\ket{\tilde\Omega}$, and keep only the first coefficients that contribute to $w_t \sim 0.999$ of the norm of $\ket{\tilde\Omega}$. This results in an important speed-up, with minimal cost, as demonstrated in Sec.~\ref{sec:further_truncation_test}. We call this new weight $w_t$ the \textit{target weight}, to distinguish it from $w_f$, even though both are weights calculated with the same sum in different state subspaces.

\section{Results}\label{sec:Results}

We now examine the validity of the method for multiple systems, exploring different sizes ($N_c$), geometries, and fractions of states ($f$). We mainly compare the resulting DOS ($A(\omega)$) and the ground state energies. 
Throughout the article, we use $t=1$ and $\eta=0.05$. We also use $U=8$ except for Fig.~\ref{fig:var_u} and $\beta = 0.2$ except for Fig.~\ref{fig:methods}. We apply periodic boundary conditions: for the linear cluster, we use a 1D periodic boundary conditions, and for the bi-dimensional cluster, we use a 2D periodic boundary conditions (with the superlattice vectors specified).
\begin{figure}[h!]
    \centering
    \includegraphics[width=\columnwidth]{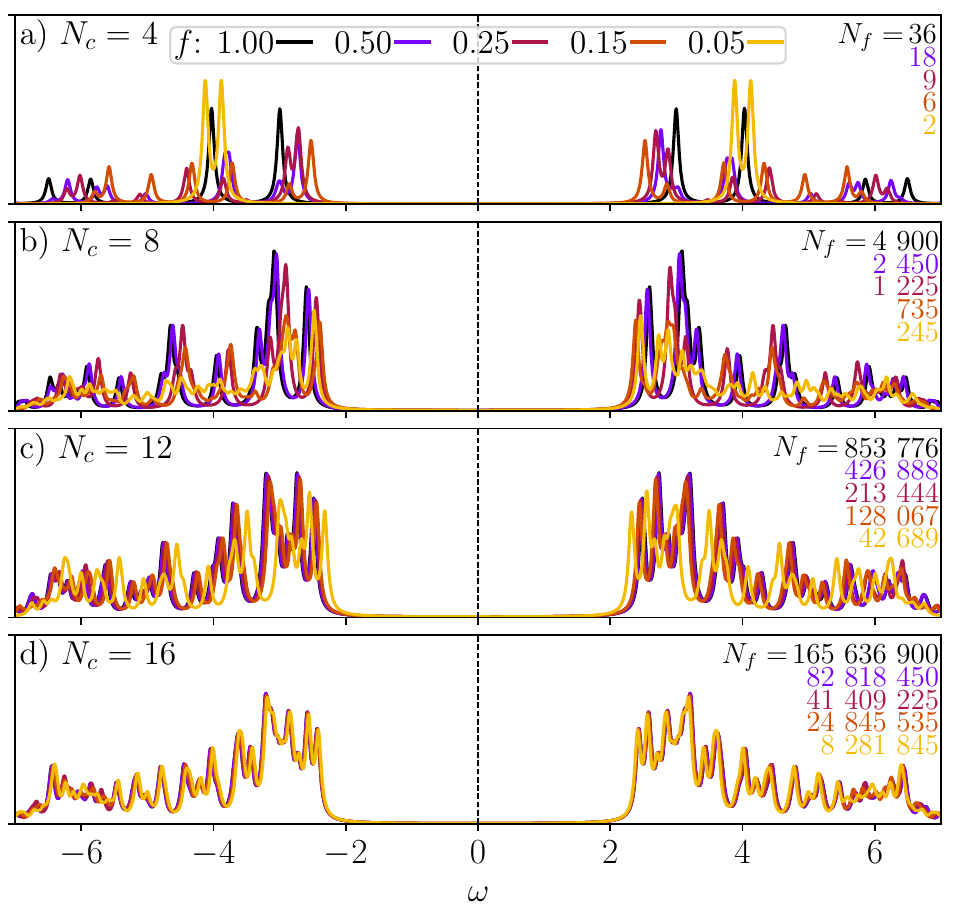}
    \vspace{-20pt}
    \caption{DOS $A(\omega)$ for multiple linear clusters: a) 4 sites, b) 8 sites, c) 12 sites, d) 16 sites. We used periodic boundary conditions in 1D. Each color represents a different fraction of states $f$, shown in the legend, of states taken according to the accessible states. On the right, the number of states associated with $f$ of the same color is shown. The reference ($f=1.0$) corresponds to normal ED. Table~\ref{tab:1D_energies} contains the corresponding ground state energy for each curve.}
    \vspace{-10pt}
    \label{fig:1D_4to16}
\end{figure}
\subsection{1D Lattice, $\mu=U/2$ half-filling}\label{sec:res_mu4}
In Fig.~\ref{fig:1D_4to16}, we compare four periodic 1D systems: 4, 8, 12, and 16 sites with half-filled systems ($N_e=N_c$, $S_z=0$, and $\mu=4$) using various fractions of states $f$. The corresponding ground state energy for each curve is shown in Table~\ref{tab:1D_energies} of Appendix~\ref{app:groundstates}. We clearly see differences between $f=1.0$ and lower fractions of states. However, these differences become much smaller as the system size increases. At 16 sites, even $f=0.05$ produces a DOS in very good agreement with the full ED ($f=1.0$).

The first effect of using fewer states is a general shift of the poles of the Green function towards the Fermi level ($\omega=0$). This is a direct consequence of the overestimated ground state energy ($\tilde\Omega > \Omega$), which shifts the poles according to Eqs.~\eqref{eq:green_qmatrix} and~\eqref{eq:green_qmatrix_m}.  
For 16 sites, the $f=0.05$ DOS is similar to the exact result at lower energies (around $\omega=0$). However, more fluctuations appear at higher energies (around $\omega\sim\pm 6$). For convenience, we use the fraction of states $f$ to compare the results, but what really matters is $N_f$. 
This result suggests that a minimum number of states is needed to accurately capture the physics, independently of the cluster size.

We can estimate the precision of the results by changing the seed of the random number generator. For well-converged cases, such as $f=0.05$ and $N_c=16$, this change of seed has no impact, unlike for $f=0.05$ and $N_c=8$. 
The standard deviation of $\tilde{\Omega}$ is $\sim10^{-1}$ for $N_c=8$ and $10^{-4}$ for $N_c=16$. The standard deviation of the gap width is $3\times10^{-1}$ and $10^{-4}$, for $N_c=8$ and $N_c=16$ respectively. 

\begin{figure}[h!]
    \centering
    \includegraphics[width=\columnwidth]{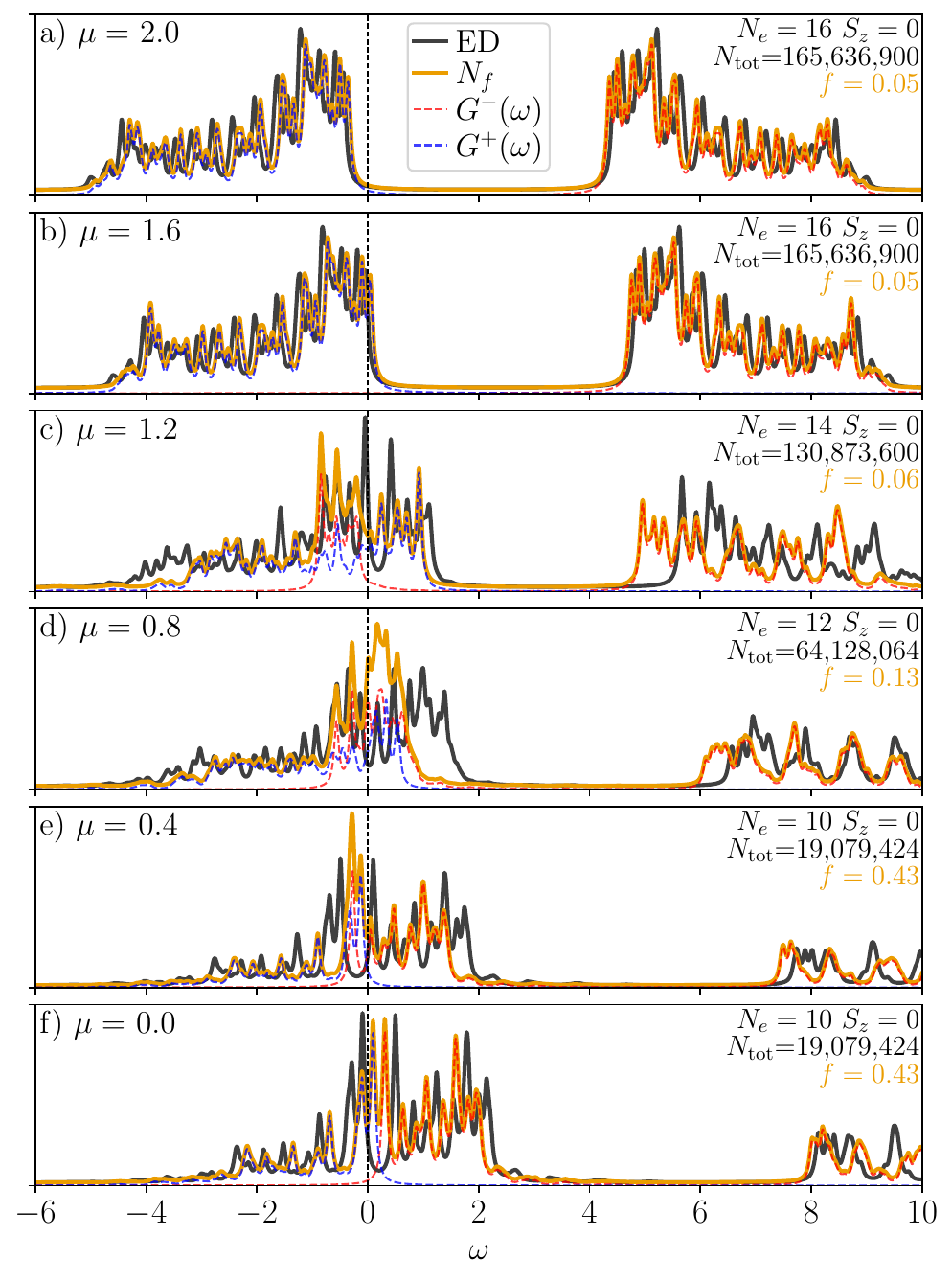}
    \vspace{-20pt}
    \caption{DOS for a 1D, 16-site lattice. In red (blue) is the hole (electron) contribution to the weight of the DOS. $N_f=8,281,845$. On the right of each plot we show the sector, its $N_\text{tot}$ and its  $f$.}
    \vspace{-20pt}
    \label{fig:1D_16_mu0}
\end{figure}

\subsection{1D Lattice, $\mu\neq U/2$}\label{sec:res_mu0}
We now check the validity of our method away from half-filling ($\mu \ne U/2$).
In Fig.~\ref{fig:1D_16_mu0}, we study the same 16-site 1D lattice presented in Fig.~\ref{fig:1D_4to16}d, while varying $\mu$, which results in changing the ground state $N_e$:$S_z$ sector.
The algorithm scales with $N_f$. If we have the ressources to use MCD for an $N_f$ at half-filling, it is also possible to use MCD in any sector, because the half-filled sector is always the largest. 
Thus we choose $N_f=8,281,845$ which corresponds to the $N_f$ of $f=0.05$ in the $N_e=16$, $S_z=0$ sector (Fig.~\ref{fig:1D_4to16}d). 



The results are good at $\mu=2.0,1.6$ and $0$, but not as much for $\mu=1.2,0.8$ and $0.4$. This disagreement comes from the error in the ground state energy.
Indeed, the $N_f$ curves are decomposed into $A^-(\omega)$ in red and $A^+(\omega)$ in blue, which are the weights from $G^-(\omega)$ and $G^+(\omega)$ (Eqs.~\eqref{eq:green_minus} and \eqref{eq:green_plus}), respectively. Normally, in any ED calculation, both of these curves should never overlap nor cross $\omega=0$, but here they do since $\tilde\Omega$ is overestimated. Apart from this effect, we observe that $A^+(\omega)$ obtained from MCD is the same as the one obtained from ED for all $\mu$ presented (same for $A^-(\omega)$). 
In principle, we could correct the overestimation of $\tilde\Omega$ by checking the highest (lowest) pole of $A^-(\omega)$ ($A^+(\omega)$) and ensuring it is below (above) $\omega=0$, applying a corrective shift to $\tilde\Omega$. But we have not yet found a robust and universal process to correct this shift.
%


\subsection{2D Lattice, half-filling}\label{sec:res_2d}
\begin{figure}[h!]
    \centering
    \includegraphics[width=\columnwidth]{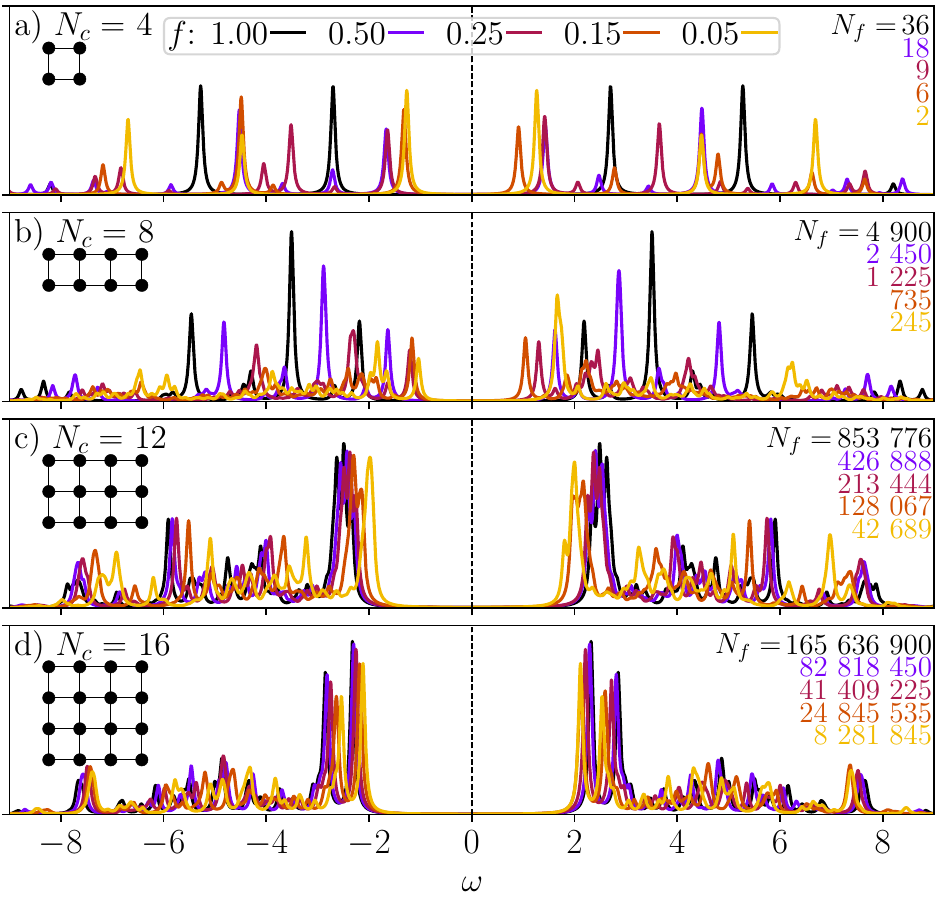}
    \caption{Same as Fig.~\ref{fig:1D_4to16} but in 2D. The cluster shapes are shown. Table~\ref{tab:2D_energies} contains the corresponding ground state energy for each curve. We used periodic boundary conditions that preserve the bipartite nature of the lattice. They are defined by 2D superlattice vectors ($\mathbf{R}_1$,$\mathbf{R}_2$) = $((2,0),(0,2))$ in (a), $((4,0),(0,2))$ in (b), $((4,0),(1,2))$ in (c) and $((4,0),(0,4))$ in (d).}
    \label{fig:2D_4sto16s}
\end{figure}

With 2D lattices, each site has twice as many neighbors as in 1D lattices. According to the analysis of Sec.~\ref{sec:ED}, we expect to require more states in $\tilde{\mathcal{F}}$ to obtain an accurate approximation. This is observed in the DOS shown in Fig.~\ref{fig:2D_4sto16s}.
At 16 sites in 2D, the $f=0.05$ approximation causes more deviation than in 1D, but is still relatively good.

We observe the same behavior in the 2D lattices as in 1D: a shift towards the Fermi level. The 16-site cluster is relatively good, but we see more noise around $\omega \sim \pm 4$ than in 1D. 
%

\subsection{1D Lattice, variable $U$}\label{sec:res_var_U}

\begin{figure}[t]
    \centering
    \includegraphics[width=\columnwidth]{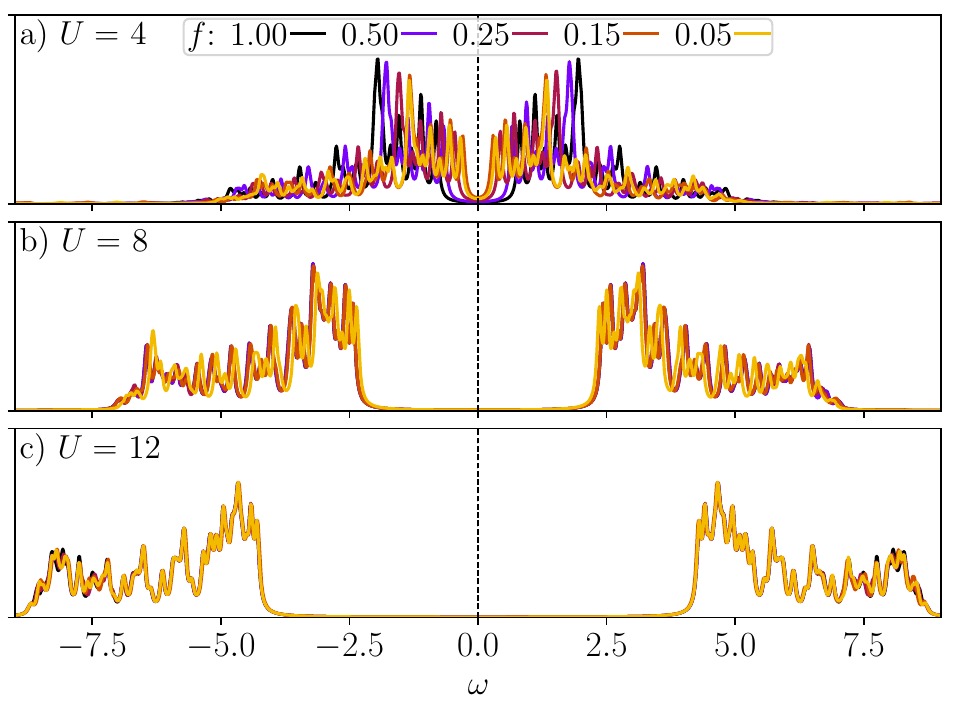}
    \caption{Same as Fig.~\ref{fig:energy_conv_1D}d, but with different $U$.}
    \label{fig:var_u}
\end{figure}

We now study the validity of MCD with different $U$. Clearly, Fig.~\ref{fig:var_u} shows that this implementation of MCD is tailored toward high-interaction problems. This is intuitive because in the low $U$ regime, the $|c_n|^2$ contributions are distributed more uniformly among all states; therefore, there are more states that are considered to be ``important''. Using a larger $U$, while keeping $\beta$ at 0.2, changes the probability of accepting a state, since $P_{a\rightarrow b}$ depends on $U\beta$. We change $\beta$ so that regardless of $U$, $P_{a\rightarrow b}$ is only caused by the change in $n_U$. $U\beta$ is set at 1.6 for all. This specific choice of $\beta$ is justified in \S\ref{sec:res_beta}.

Furthermore, we note that the MCD method formulated in the Fourier space should have more success for low $U$, but this is reserved for later work.

\subsection{Other sampling methods and $\beta$}\label{sec:res_beta}
\begin{figure}[h]
    \centering
    \includegraphics[width=\columnwidth]{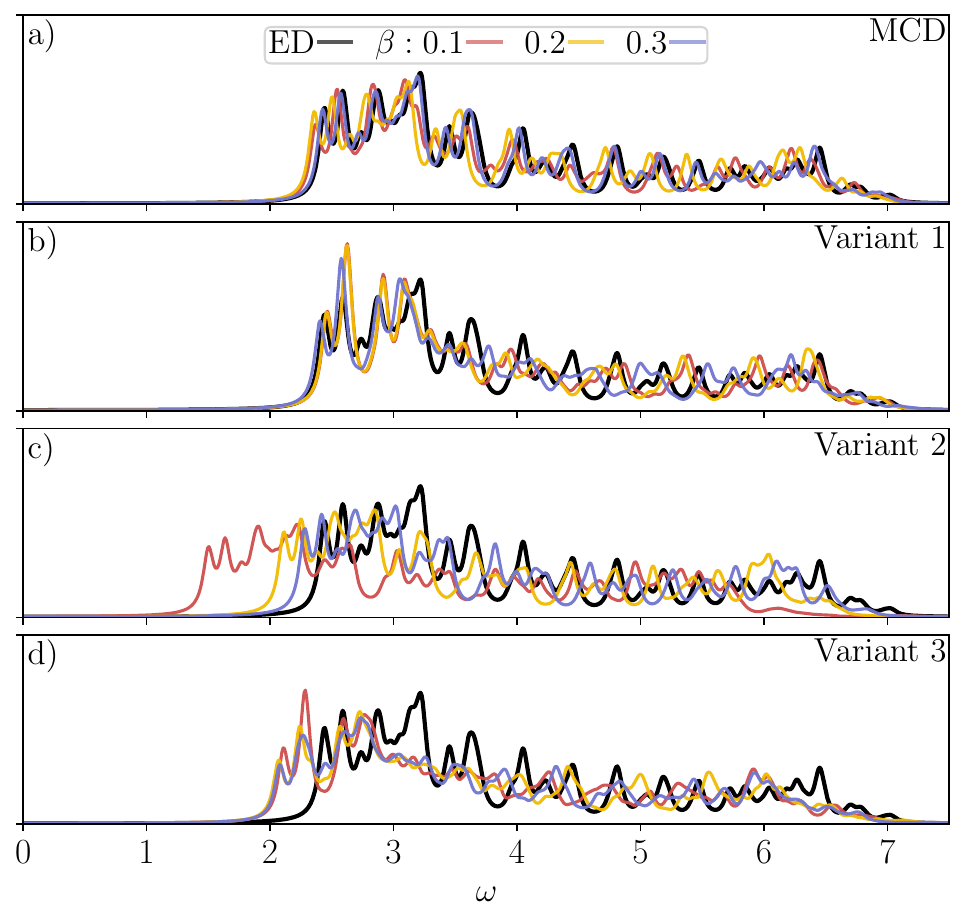}
    \vspace{-20pt}
    \caption{Presentation of MCD and the three sampling variants for different values of $\beta$. The colored lines represent an $f=0.05$ sampling for different values of $\beta=0.1,0.2,0.3$. The system is a linear 16-site periodic cluster. Table~\ref{tab:methods_energies} contains the corresponding ground state energy for each curve.}
    \label{fig:methods}
\end{figure}

Before settling on the samping process presented in \S\ref{sec:sampling}, multiple variants have been tested. They are all compared in Fig.~\ref{fig:methods}. The variants are:
\begin{enumerate} 
    \item We can use a ratio of Boltzmann factors when deciding if the proposed state should be added to $\tilde{\mathcal{F}}$:
    \begin{equation}\label{eq:boltzmann_factor2}
    \hspace{1cm}P^{\text{var 1}}_{a \rightarrow b} = \min\Big(\exp\big({-U\beta ({n_U}_b-{n_U}_a})\big),1\Big).
    \end{equation}
    %
    \item Instead of applying every possible hopping to each newly added state, we apply only one hopping generated randomly. It is accepted with $P_{a\rightarrow b}$ and only one state will act as a ``newly added''. This changes the behavior of the sampling which now is more similar to a depth-first search. 
    \item Lastly, we completely change the process of proposing states. Instead of rellying on the ``newly added'' states, we choose a random state in the growing subspace $\tilde{\mathcal F}$ and apply one random hopping. We accept the suggested state with $P_{a\rightarrow b}$. This does not favor the connectivity of the states in $\tilde{\mathcal F}$ and therefore much information is lost. 
\end{enumerate}
It is clear from these results that MCD generates better results.
For MCD and its variants, we tested multiple $\beta$. MCD and variant 2 become better when $\beta$ is increased. Variants 1 and 3 are not as $\beta$ dependent as the other two. 

If we focus on MCD, increasing the $\beta$ gives better results, but takes more time. A $\beta=0.3$ sampling takes $\sim$300 times longer than a $\beta=0.2$, which samples for $\sim$10 minutes.


\subsection{Further truncation test}\label{sec:further_truncation_test}

\begin{figure}[h]
    \centering
    \includegraphics[width=\columnwidth]{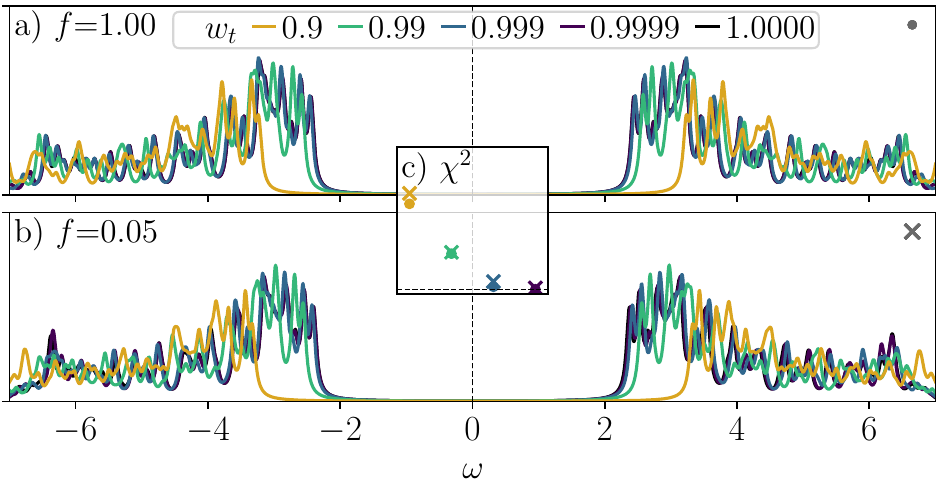}
    \caption{DOS comparison for different $f$ and $w_t$ combinations, for the linear 1D cluster. We compare a) $f=1.00$ ($\bullet$) and b) $f=0.05$ ($\times$). The same combinations of $f$ and $w_t$ are presented in Table~\ref{tab:reduce_green}, showing the size of the corresponding subspaces. c) $\chi^2$ value between the exact $A_{1,1}(\omega)$ (at $f=w_t=1$) and the truncated $A_{f,w_t}(\omega)$.}
    \label{fig:reduce_green}
\end{figure}

In Sec.~\ref{sec:truncation}, we discussed a last truncation that we can do in order to minimize the calculation time of the Green function with almost no cost. In Fig.~\ref{fig:reduce_green}, together with Table~\ref{tab:reduce_green}, we show the effect and the basis reduction associated with this truncation on the 16-site cluster in 1D at half-filling by measuring the error, estimated by:
\begin{align}\label{eq:chi2}
\chi^2 = \int d\omega \,\big(A_{f,w_t}(\omega) - A_{1,1}(\omega)\big)^2.
\end{align}
It is clear from Fig.~\ref{fig:reduce_green} that if $w_t \sim 0.999$, essentially no information is lost, as the error is essentially null. In fact, the $w_t=0.999$ and $w_t=1.0$ curves are almost indistinguishable except at higher energies (around $\omega\sim\pm 7$). We show both results for $f=1.0$ and $f=0.05$ to confirm that the behavior is independent of $f$.

Table~\ref{tab:reduce_green} shows the number of states associated with the different combinations of $f$ and $w_t$. We see that for $f=0.05$, the $\tilde{\mathcal{F}}^\pm$ size is always larger than $\tilde{\mathcal{F}}$. Most importantly, at $f=0.05$, the $\tilde{\mathcal{F}}$ size at $w_t=0.999$ is already 10 times smaller than at $w_t=1.0$. For $w_t=0.999$, the $\tilde{\mathcal{F}}^\pm$ size increases compared to $\tilde{\mathcal{F}}$, but remains smaller than the size of $\tilde{\mathcal{F}}$ at $w_t=1$. Hence, with this truncation, we remove the bottleneck on the Green function representation with minimal loss of information.

\begin{table}[h]
\setlength{\tabcolsep}{5pt}
\begin{tabular}{c|lrr} 
\toprule
$f$ & \quad$w_t$ \quad  &  \quad\quad\quad$\tilde{\mathcal{F}}$ size \quad & \quad\quad\quad$\tilde{\mathcal{F}}^\pm$ size \quad\\
\midrule
\multirow{5}{*}{\centering
\begin{tabular}{@{}c@{}}
1.00 \\ ($\bullet$)
\end{tabular}
} 
& 1.0000  &     165,636,900  & \phantom{$\sim$}147,232,800\\ 
& 0.9999  & $\sim$6,813,000  & $\sim$24,328,000 \\
& 0.999   & $\sim$2,285,000  & $\sim$10,085,000 \\ 
& 0.99    &   $\sim$483,000  &  $\sim$2,616,000 \\ 
& 0.9     &    $\sim$41,000  &    $\sim$265,000 \\ 
\midrule
\multirow{5}{*}{\centering
\begin{tabular}{@{}c@{}}
0.05 \\ ($\times$)
\end{tabular}
} 
& 1.0000  & \phantom{$\sim$}8,281,845  & $\sim$23,000,000 \\
& 0.9999  &           $\sim$1,603,000  &  $\sim$7,700,000 \\
& 0.999   &             $\sim$806,000  &  $\sim$4,200,000 \\
& 0.99    &             $\sim$256,000  &  $\sim$1,493,000 \\
& 0.9     &              $\sim$28,000  &    $\sim$192,000 \\
\bottomrule
\end{tabular}
    \caption{Size of the subspaces $\tilde{\mathcal{F}}$ and $\tilde{\mathcal{F}}^\pm$ associated with the same combinations of $f$ and $w_t$ as in Fig.~\ref{fig:reduce_green}.
    The approximate values depend on the random number generator seed.}
    \label{tab:reduce_green}
\end{table}
%

\subsection{Large Clusters}\label{sec:24_site}

\begin{figure}[h]
    \centering
    \includegraphics[width=\columnwidth]{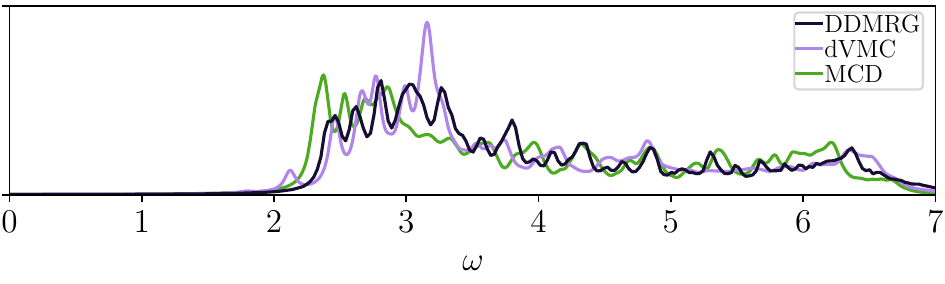}
    \vspace{-20pt}
    \caption{DOS obtained from DDMRG, dVMC and MCD. Here, the system is a linear 1D cluster with $N_c=24$ for $N_f=75\times 10^6$ and $w_t=0.999$.}
    \label{fig:24_sites}
\end{figure}

Compared to ED, the MCD method is faster and requires less computer memory. With that in mind, we push the cluster size beyond what is computable with ED. We see, in Fig.~\ref{fig:24_sites}, a 1D 24-site cluster. We compare MCD with the dynamical extension of DMRG (called DDMRG) and of VMC (called dVMC). DDMRG was computed with the DDMRG++ algorithm~\cite{10.1021/acs.jctc.7b00682} from the package \texttt{pyblock3}~\cite{pyblock3}. The dVMC used the package published in Ref~\cite{Charlebois2020}. The MCD computation has been done with a fixed $N_f=75\times10^6$ and $w_t=0.999$. 
DDMRG, in 1D, might as well be considered exact, but is rather slow compared to MCD as it requires a DMRG solution for every $\omega$ computed.

We obtain the expected physics. The Mott gap is well defined, even if slightly underestimated compared to the DDMRG result. The overall bandwidth is comparable to the DDMRG and dVMC results. With MCD, we obtained $\tilde \Omega/N_c=$ -4.2714, while DDMRG and dVMC resulted in a $\Omega/N_c$ of -4.3282 and -4.3250, respectively.
The dVMC method is shown to compare MCD with another Monte Carlo method, with comparable sampling time of the Fock space.

\section{Discussion and conclusion}\label{sec:discussion}
The results show that the ground state and Green function can effectively be approximated by a subset of states of the full Fock space. 
The main hurdle remains the overestimation of the ground state energy, which results in a slight translation of both the electron and hole Green functions, and thus an underestimation of any gap near the Fermi level. 


The MCD method shares similarities with other truncated vector space methods. However, the nature of the vector space is very different. In QMCD, the states are Slater determinants of different configurations of auxiliary fields. QMCD is very close to QMC, where an extra step of diagonalization is added in order to remove the sign problem. Whereas MCD is very close to ED, where an extra structure (namely a binary tree in this article) is kept and sampled in order to reduce the size of the Fock space. 

As for the DDMRG, the idea is also to find the minimal representation of the physics. However, the subspace generated with the matrix product states is so small and optimized for the matrix product object that one cannot compute the complete DOS in one go. A new DMRG solution and subspace need to be calculated for each frequency $\omega$ of the DOS. It is also to be noted that matrix product states are conceptually very far from the Fock basis.

MCD is really close to ED and can be implemented easily in any existing ED solver. A first implementation that saves time and memory is to truncate the ground state subspace for the Green function calculation (see Fig.~\ref{fig:reduce_green} and Table~\ref{tab:reduce_green}) without any cost on the precision for $w_t=0.999$. Also, a significant gain can be obtained in speed already with $f=0.50$, which is essentially exact even in 2D. This increase in efficiency opens the door to using large clusters in embedding techniques like CDMFT. 

Let us try to better understand the idea behind the MCD sampling \textit{à la BFS}. The Hubbard model has two competing energies: the kinetic part (negative), which has a tendency to delocalize the electrons, and the interaction part (positive), which has a tendency to localize the electrons. 
The ``most important states'' generally have lower double occupation but also more ``hopping possibilities''. From the point of view of graph theory, the more links there are to a specific Fock state $\ket{\tilde{x}_n}$, the more probable this state is. In our test to propose new states, the Boltzmann weight in Eq.~\eqref{eq:boltzmann_factor} only accounts directly for the interaction energy $U$ and not for the number of states connected to the new state. However, the probability of suggesting states with higher connectivity is greater, since there are more paths to reach them. 

From the graph point of view, if the graph of $\tilde{\mathcal{F}}$ is separated into two or more parts, it means that the associated Hamiltonian matrix $\mathbf{\tilde H}$ is composed of two or more blocks. The ground state will be in only one of the blocks, meaning that all the other blocks contain states that do not contribute to the ground state.
On the other hand, a more connected graph results in a lower ground state energy. This is why a search algorithm \textit{à la} breadth-first search will result in a lower $\Omega$. Note that some high-$n_U$ states are less ``important'' in the sense that their $|c_n|^2$ is much lower, but keeping them is still important to find the ground state, 
because they are responsible for connecting different states. As an extreme example, the two AFM states ($n_U=0$) are always the ``most important'' in the half-filled sector, but they are extremely far from each other. In fact, they are disconnected from any other $n_U=0$ states and require hopping to an $n_U=1$ state before reaching any other part of the graph. Usually, for smaller systems, the $n_U=0$ level fills quickly, ensuring that both antiferromagnetic states are present and connected. However, for the 24-site clusters in 1D, seeding the search with only one of the AFM states does not guarantee reaching the second AFM state. If, on the other hand, we seed the search with both AFM states, this will result at first in two disconnected graphs that will eventually connect to each other. However, the ground state energy is worse, and the symmetry of the DOS and the finite-size effect are lost. This is due to the lower connectivity of the resulting graph. This graph connectivity is further explained in Appendix~\ref{app:graph} with a 4-site cluster.

It would be interesting to study the connectivity of the resulting graph of different clusters and geometries in future work, to better understand the most probable states. It would also be interesting to explore different bases and to study the impact of symmetries. Indeed, more symmetries result in smaller sectors 
and $\mathcal{F}^\pm$ contain states excited from sectors that cannot result from Eq.~\eqref{eq:excited_subspace}. It will be necessary to apply $\hat H$ to the most important states of $\tilde{\mathcal{F}}^\pm$ in order to obtain better results. A detailed treatment is left for future work.

\begin{acknowledgments}
We acknowledge useful discussions with D. Sénéchal. The original code developed for this project has been benchmarked with the help of the ED impurity solver of PyQCM~\cite{pyqcm}.
This project was supported by the Natural Sciences and Engineering Research Council of Canada (NSERC) under Grant No. RGPIN-2021-04043. B.B. was supported by a scholarship from the Fonds de recherche du Québec – Nature et technologies (FRQNT).
\end{acknowledgments}

\appendix
\section{Ground state energy}
\label{app:groundstates}
Here we tabulate all the ground state energies per site of the solutions presented in \S\ref{sec:Results}.
\begin{table}[h!]
\setlength{\tabcolsep}{5pt}
\begin{tabular}{c|rrrrr} 
\toprule
 $N_c$ & $f=1.00$ & $f=0.50$ & $f=0.25$ & $f=0.15$ & $f=0.05$\\
\midrule
 4 & -4.3300 & -4.2862 & -4.1383 & -4.0000 & -4.0307\\
 8 & -4.3332 & -4.3228 & -4.2863 & -4.3019 & -4.2586\\
 12 & -4.3302 & -4.3267 & -4.3252 & -4.3179 & -4.2672\\
 16 & -4.3290 & -4.3282 & -4.3270 & -4.3167 & -4.3187\\
\bottomrule
\end{tabular}
\caption{Values of $\tilde\Omega/N_c$ in 1D, for each curve presented in Fig.~\ref{fig:1D_4to16}.}
\label{tab:1D_energies}
\vspace{-10pt}
\end{table}
\begin{table}[h!]
\setlength{\tabcolsep}{5pt}
\begin{tabular}{c|rr}
\toprule
 $\mu$& ED & MCD \\
\midrule
$2.0$ & -2.3290 & -2.3196 \\
$1.6$ & -1.9290 & -1.9196 \\
$1.2$ & -1.5776 & -1.4900 \\
$0.8$ & -1.2605 & -1.2092 \\
$0.4$ & -0.9829 & -0.9589 \\
$0.0$ & -0.7329 & -0.7206 \\
\bottomrule
\end{tabular}
\caption{Values of $\tilde\Omega/N_c$ for each curve presented in Fig.~\ref{fig:1D_16_mu0}.}
\label{tab:mu_energies}
\vspace{-10pt}
\end{table}
\begin{table}[h!]
\setlength{\tabcolsep}{5pt}
\begin{tabular}{c|rrrrr} 
\toprule
 $N_c$ & $f=1.00$ & $f=0.50$ & $f=0.25$ & $f=0.15$ & $f=0.05$\\
\midrule
4 & -4.3300 & -4.2862 & -4.1383 & -4.0000 & -4.0307\\ 
8 &  -4.8908 & -4.8020 &-4.6961 &-4.7344 &-4.5852\\ 
12 & -4.5226 &-4.5127 &-4.5032 &-4.4658 &-4.4042\\ 
16 & -4.5293 &-4.5263 &-4.5180 &-4.5027 &-4.4892\\
\bottomrule
\end{tabular}
\caption{Values of $\tilde\Omega/N_c$ in 2D, for each curve presented in Fig.~\ref{fig:2D_4sto16s}.}
\label{tab:2D_energies}
\vspace{-10pt}
\end{table}
\begin{table}[h!]
\setlength{\tabcolsep}{5pt}
\begin{tabular}{c|rrr} 
\toprule
 Methods & $\beta=0.1$ & $\beta=0.2$ & $\beta=0.3$\\
\midrule
MCD & -4.3237 & -4.3196 & -4.3266 \\ 
Variant 1 &  -4.3211 & -4.3213 &-4.3204\\ 
Variant 2 & -4.2512 &-4.2963 &-4.3072\\ 
Variant 3 & -4.2957 &-4.2943 &-4.2944\\
\bottomrule
\end{tabular}
\caption{Values of $\tilde\Omega/N_c$ for each curve presented in Fig.~\ref{fig:methods}. The ED result is $\Omega/N_c=-4.3290$}.
\label{tab:methods_energies}
\vspace{-10pt}
\end{table}
\begin{table}[h!]
\setlength{\tabcolsep}{5pt}
\begin{tabular}{c|rrrrr} 
\toprule
 $U$ & $f=1.00$ & $f=0.50$ & $f=0.25$ & $f=0.15$ & $f=0.05$\\
\midrule
 4 & -2.5759 & -2.5629 & -2.5412 & -2.5190 & -2.5056\\
 8 & -4.3332 & -4.3228 & -4.2863 & -4.3019 & -4.2586\\
 12& -6.2261 & -6.2261 & -6.2261 & -6.2248 & -6.2257\\
\bottomrule
\end{tabular}
\caption{Values of $\tilde\Omega/N_c$ in 1D, for each curve presented in Fig.~\ref{fig:var_u}.}
\label{tab:var_U_energies}
\end{table}

\section{Lanczos Algorithm}
\label{app:lanczos_algorithm}
The Lanczos algorithm~\cite{lanczos_iteration_1950,Caffarel:1994,dagotto,bai_templates_2000} is a simple and time-efficient way to compute the ground state $\ket \Omega$ and its energy $\Omega$  
of the Hamiltonian $\hat H$. Note that we could work with the matrix $\mathbf H$ directly, but to avoid storing this very large matrix in memory, we apply the operator $\hat H$ instead. 
This is an iterative process where we construct an orthonormal basis in the Krylov subspace:
\begin{equation}\label{app::eq::krylov}
 \mathcal{K}_r = \qty{\ket{\phi_0},\hat H\ket{\phi_0},\hat H^2\ket{\phi_0},\dots,\hat H^{r}\ket{\phi_0}}
\end{equation}
where $\ket{\phi_0}$ is a random initial vector.
The orthogonal basis is generated using the recursion relation:
\begin{equation}\label{app::eq::next_vector}
 \ket{\phi_{n+1}} = \hat{H}\ket{\phi_{n}} - a_n\ket{\phi_{n}}- b_n\ket{\phi_{n-1}},
\end{equation}
where:
\begin{align}\label{app::eq::an_bn}
 a_n = \frac{\ev{\hat H}{\phi_n}}{\ip{\phi_n}} && b_n^2 = \frac{\ip{\phi_n}}{\ip{\phi_{n-1}}},
\end{align}
and where $b_0=0$ and $\ket{\phi_{-1}}=0$.
The basis $\ket{ \phi_n}$ is then normalized as $\ket{n} = \ket{ \phi_n}/\braket{\phi_n}{\phi_n}$. In this subspace, $\hat H$ can be expressed as an effective representation of the Hamiltonian with $H^{\rm{eff}}_{nm}=\bra{n}{\hat H}\ket{m}$, resulting in a tridiagonal matrix:
\begin{equation}
\mathbf{H}_{\rm{eff}}=\left(\begin{array}{ccccc}
a_0 & b_1 &     &  \\
b_1 & a_1 & b_2 &  \\
  & b_2 & a_2 & \scriptsize{\ddotss} \\
  &     & \scriptsize{\ddotss} & \scriptsize{\ddotss} \\
\end{array}\right)
\end{equation}
The ground state $\ket{\Omega_{\rm{eff}}}$ and ground state energy $\Omega_{\rm{eff}}$ of this effective matrix $\mathbf{H}_{\rm{eff}}$ are good approximations of $\ket{\Omega}$ and $\Omega$.
With each iteration, $\mathbf{H}_{\rm{eff}}$ grows and so does the accuracy of the ground state. This converges in a few hundred iterations. 
The ground state of $\mathbf H$ in the full Fock space can be reconstructed by calculating:
\begin{equation}\label{app::eq::change_basis}
 \ket{\Omega} = \sum_n\dyad{n} \Omega_{\rm{eff}}\rangle,
\end{equation}
which is an important step to prepare for the band Lanczos algorithm.

Up to this point, the Lanczos algorithm is the same for traditional ED and Monte Carlo diagonalization. We just need to replace $\mathbf{H}_{\rm{eff}}$ with $\tilde{\mathbf{H}}_{\rm{eff}}$ (and every resulting quantity like $|\tilde{\Omega}_{\rm{eff}}\rangle$, $|\Omega_{\rm{eff}}\rangle$). However, in Monte Carlo diagonalization, the operation $\hat H\ket{\phi_{n}}$ will generate Fock states $|x\rangle$ that are outside of $\tilde{\mathcal{F}}$.
These emerging states are simply discarded. Discarding those Fock states has the same effect as constructing the matrix $\mathbf{\tilde H}$ in the subspace and applying it to the vector.

\section{Band Lanczos Algorithm}
\label{app:band_lanczos_algorithm}
The Band Lanczos algorithm~\cite{Caffarel:1994,dagotto,bai_templates_2000} is a generalization of the Lanczos algorithm. Closely clustered eigenvalues and reduced-order modeling favor the use of this algorithm.  
Unlike the Lanczos algorithm, this method uses $L$ initial vectors $\ket{\phi_1},\dots,\ket{\phi_L}$ and creates an orthonormal basis in the Krylov subspace $\mathcal{K}_r^L$:
\begin{equation}
\begin{aligned}
    \mathcal{K}_r^L=\{&\ket{\phi_1},\dots,\ket{\phi_L},\\
    &\hat H\ket{\phi_1},\dots,\hat H\ket{\phi_L},\dots,\\
    &\hat H^{r}\ket{\phi_1},\dots,\hat H^{r}\ket{\phi_L}\}
\end{aligned}
\end{equation}
The algorithm is well suited to calculate the electron Green function $G^{+}_{\mu\nu}(\omega)$ because it has $N_c$ important vectors $\hat{c}_{\nu}^\dag|\Omega\rangle$. Thus, we choose $\ket{\phi_\nu} = \hat{c}_{\nu}^\dag|\Omega\rangle$ as the $L=N_c$ starting vectors to generate an orthonormal basis in which we can express the effective Hamiltonian of the excited subspace $\mathbf{H}_{\rm{eff}}^+$. The eigenvalues and eigenvectors of this resulting matrix are $E_n^+$ and $|E_n^+\rangle$. For the Green function, every eigenpair is necessary. For the hole Green function $G^{-}_{\mu\nu}(\omega)$, we can perform another band Lanczos with $\ket{\phi_\nu} = \hat{c}_{\nu}|\Omega\rangle$, resulting in $E_n^-$ and $|E_n^-\rangle$.

Up to this point, the band Lanczos algorithm is the same for traditional ED and Monte Carlo diagonalization. We just need to replace $\mathbf{H}_{\rm{eff}}^\pm$ with $\tilde{\mathbf{H}}_{\rm{eff}}^\pm$ and $\ket{\phi_\nu} = \hat{c}_{\nu}^{(\dag)}|\Omega\rangle$
with 
$|\tilde{\phi}_\nu\rangle = \hat{c}_{\nu}^{(\dag)}|\tilde{\Omega}\rangle$
(and every resulting quantity like $\tilde{E}_{n}^\pm$, $|\tilde{E}_{n}^\pm\rangle$). However, in Monte Carlo diagonalization, the operation $\hat H\ket{\phi_{n}}$ will generate Fock states $|x\rangle$ that are outside of $\tilde{\mathcal{F}}^\pm$. These emerging states are also discarded.

\section{BFS vs DFS}\label{app:graph}
\begin{figure}[H]
    \centering
    \vspace{1em}
\includegraphics[width=0.99\linewidth]{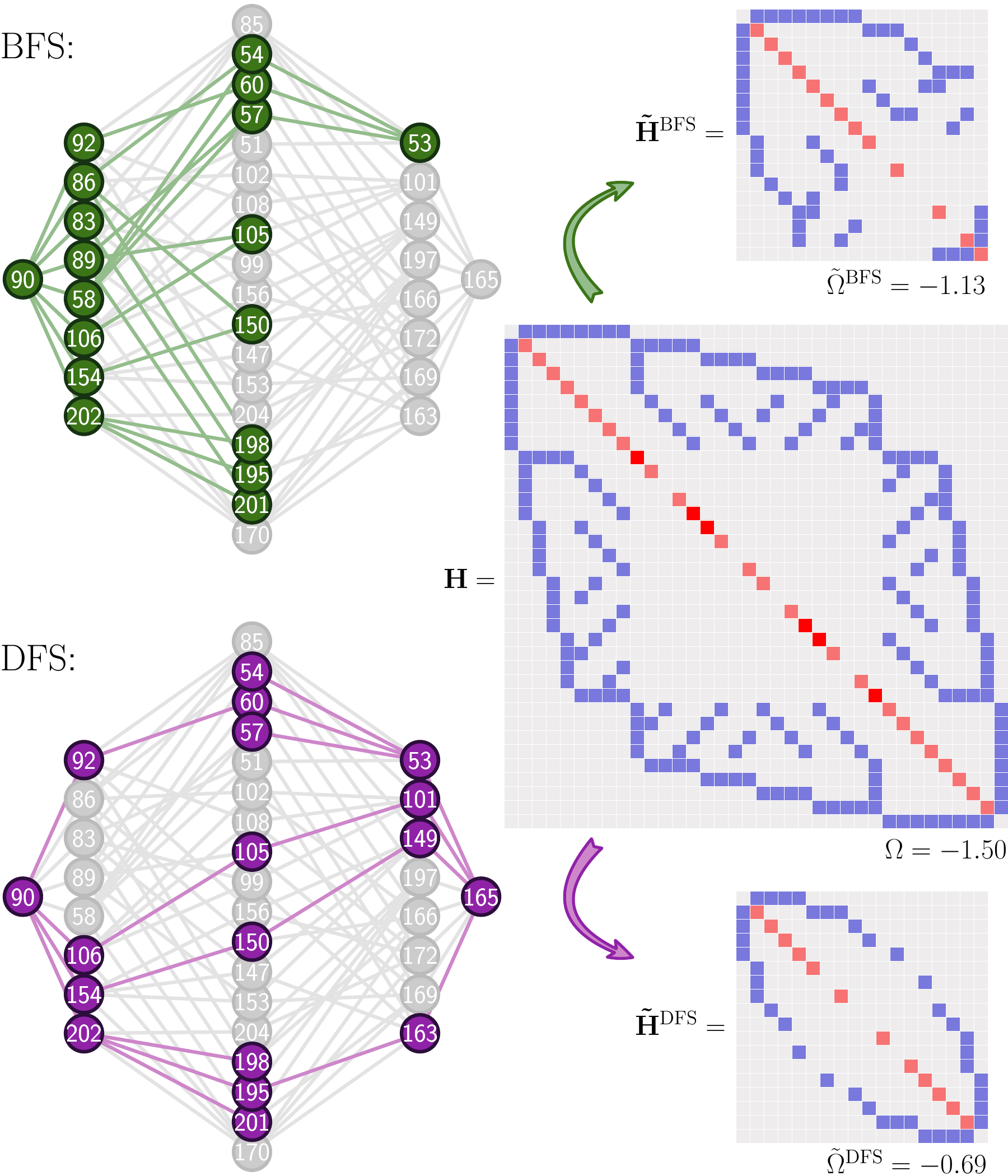}
\caption{A 4-site linear cluster Hamiltonian and its approximations. The $36\times36$ pixel matrix $\mathbf{H}$ is the full Hamiltonian sector where $N_e=4$ and $S_z=0$. The blue squares represent the terms $-t=-1$, the gray squares are the zeros and the red squares are multiples of $U=8$. The BFS graph shows the 18 states selected
through a proposition scheme \textit{à la BFS} (top-left) and
the resulting $\mathbf{\tilde{H}}^{\mathrm{BFS}}$ matrix (top-right). The DFS graph shows the 18 states selected
through a proposition scheme \textit{à la DFS} (bottom-left) and
the resulting $\mathbf{\tilde{H}}^{\mathrm{DFS}}$ matrix (bottom-right).}
\vspace{-10pt}
    \label{fig:graphs}
\end{figure}

Here we present the 4-site cluster graph of all the states with
$N_e=4$ and $S_z=0$. The results are summarized in  Fig.~\ref{fig:graphs}. The full Hamiltonian in this sector, $\mathbf{H}$, is a 
$36\times36$ matrix. 
If we propose the states with a BFS approach,
as it is done in MCD, we obtain a more connected graph, with a
sub-matrix $\mathbf{\tilde{H}}^{\mathrm{BFS}}$ containing more
negative terms (blue pixels). Otherwise, if we propose the states with a DFS approach, as is done n Fig.~\ref{fig:methods}c (variant 2), the graph is less connected and $\mathbf{\tilde{H}}^{\mathrm{DFS}}$ contains fewer negative terms. Although both approaches are inexact, the ground state energy is much closer to the exact results for a BFS sampling than for a DFS sampling.

To understand the state labels of Fig.~\ref{fig:graphs}, we used the convention that binary numbers represent the occupation. For example, the state:
\begin{equation}
\ket{90} = \vert\underbrace{0101}_{\text{spin}\uparrow };\underbrace{1010}_{\text{spin}\downarrow }\rangle
\end{equation}
represents one of the antiferromagnetic states, the other one being $\ket{165}$. Both antiferromagnetic states are at each end of the graphs. Even if they have an important contribution to the ground state, a subspace including both will not result in a better representation of the ground state, as it is evident from the ground state energies of Fig.~\ref{fig:graphs}.

\bibliography{refs}

@article{cdmft-kotliar,
    title = {Cellular Dynamical Mean Field Approach to Strongly Correlated Systems},
    author = {Kotliar, Gabriel and Savrasov, Sergej Y. and P\'alsson, Gunnar and Biroli, Giulio},
    journal = {Phys. Rev. Lett.},
    volume = {87},
    issue = {18},
    pages = {186401},
    numpages = {4},
    year = {2001},
    month = {Oct},
    publisher = {American Physical Society},
    doi = {10.1103/PhysRevLett.87.186401},
    url = {https://link.aps.org/doi/10.1103/PhysRevLett.87.186401}
}

@article{cdmft-lichtenstein,
    title = {Antiferromagnetism and d-wave superconductivity in cuprates: A cluster dynamical mean-field theory},
    author = {Lichtenstein, A. I. and Katsnelson, M. I.},
    journal = {Phys. Rev. B},
    volume = {62},
    issue = {14},
    pages = {R9283--R9286},
    numpages = {0},
    year = {2000},
    month = {Oct},
    publisher = {American Physical Society},
    doi = {10.1103/PhysRevB.62.R9283},
    url = {https://link.aps.org/doi/10.1103/PhysRevB.62.R9283}
}

@article{pyqcm,
    author={Dionne, Théo N. and Foley, Alexandre and Rousseau, Moïse and Sénéchal, David},
    title={{Pyqcm: An open-source Python library for quantum cluster methods}},
    journal={SciPost Phys. Codebases},
    pages={23},
    year={2023},
    publisher={SciPost},
    doi={10.21468/SciPostPhysCodeb.23},
    url={https://scipost.org/10.21468/SciPostPhysCodeb.23}
}

@article{hubbard-hamiltonian,
    author = {Hubbard, J.  and Flowers, Brian Hilton },
    title = {Electron correlations in narrow energy bands},
    journal = {Proceedings of the Royal Society of London. Series A. Mathematical and Physical Sciences},
    volume = {276},
    number = {1365},
    pages = {238-257},
    year = {1963},
    doi = {10.1098/rspa.1963.0204},
    
    URL = {https://royalsocietypublishing.org/doi/abs/10.1098/rspa.1963.0204},
}

@article{dagotto,
    title = {Correlated electrons in high-temperature superconductors},
    author = {Dagotto, Elbio},
    journal = {Rev. Mod. Phys.},
    volume = {66},
    issue = {3},
    pages = {763--840},
    numpages = {0},
    year = {1994},
    month = {Jul},
    publisher = {American Physical Society},
    doi = {10.1103/RevModPhys.66.763},
    url = {https://link.aps.org/doi/10.1103/RevModPhys.66.763}
}

@article{VMC2008,
author = {Tahara ,Daisuke and Imada ,Masatoshi},
title = {Variational Monte Carlo Method Combined with Quantum-Number Projection and Multi-Variable Optimization},
journal = {Journal of the Physical Society of Japan},
volume = {77},
number = {11},
pages = {114701},
year = {2008},
doi = {10.1143/JPSJ.77.114701},
URL = {https://doi.org/10.1143/JPSJ.77.114701}
}

@article{Charlebois2020,
  title = {Single-{{Particle Spectral Function Formulated}} and {{Calculated}} by {{Variational Monte Carlo Method}} with {{Application}} to \$d\$-{{Wave Superconducting State}}},
  author = {Charlebois, Maxime and Imada, Masatoshi},
  year = {2020},
  month = nov,
  journal = {Physical Review X},
  volume = {10},
  number = {4},
  pages = {041023},
  publisher = {{American Physical Society}},
  doi = {10.1103/PhysRevX.10.041023},
}

@article{Rosenberg2022,
  title = {Fermi arcs from dynamical variational Monte Carlo},
  author = {Rosenberg, P. and S\'en\'echal, D. and Tremblay, A.-M. S. and Charlebois, M.},
  journal = {Phys. Rev. B},
  volume = {106},
  issue = {24},
  pages = {245132},
  numpages = {13},
  year = {2022},
  month = {Dec},
  publisher = {American Physical Society},
  doi = {10.1103/PhysRevB.106.245132},
  url = {https://link.aps.org/doi/10.1103/PhysRevB.106.245132}
}

@Article{Qin_Schafer_Andergassen_Corboz_Gull_2022,
  author       = {Qin, Mingpu and Schäfer, Thomas and Andergassen, Sabine and Corboz, Philippe and Gull, Emanuel},
  journal      = {Annual Review of Condensed Matter Physics},
  title        = {The Hubbard Model: A Computational Perspective},
  year         = {2022},
  number       = {1},
  pages        = {275–302},
  volume       = {13},
  doi          = {10.1146/annurev-conmatphys-090921-033948},
}

@article{LeBlanc2015,
  title = {Solutions of the Two-Dimensional Hubbard Model: Benchmarks and Results from a Wide Range of Numerical Algorithms},
  author = {LeBlanc, J. P. F. and Antipov, Andrey E. and Becca, Federico and Bulik, Ireneusz W. and Chan, Garnet Kin-Lic and Chung, Chia-Min and Deng, Youjin and Ferrero, Michel and Henderson, Thomas M. and Jim\'enez-Hoyos, Carlos A. and Kozik, E. and Liu, Xuan-Wen and Millis, Andrew J. and Prokof'ev, N. V. and Qin, Mingpu and Scuseria, Gustavo E. and Shi, Hao and Svistunov, B. V. and Tocchio, Luca F. and Tupitsyn, I. S. and White, Steven R. and Zhang, Shiwei and Zheng, Bo-Xiao and Zhu, Zhenyue and Gull, Emanuel},
  collaboration = {Simons Collaboration on the Many-Electron Problem},
  journal = {Phys. Rev. X},
  volume = {5},
  issue = {4},
  pages = {041041},
  numpages = {28},
  year = {2015},
  month = {Dec},
  publisher = {American Physical Society},
  doi = {10.1103/PhysRevX.5.041041},
  url = {https://link.aps.org/doi/10.1103/PhysRevX.5.041041}
}

@article{Schaefer2021,
  title = {Tracking the Footprints of Spin Fluctuations: A MultiMethod, MultiMessenger Study of the Two-Dimensional Hubbard Model},
  author = {Sch\"afer, Thomas and Wentzell, Nils and \ifmmode \check{S}\else \v{S}\fi{}imkovic, Fedor and He, Yuan-Yao and Hille, Cornelia and Klett, Marcel and Eckhardt, Christian J. and Arzhang, Behnam and Harkov, Viktor and Le R\'egent, Fran\ifmmode \mbox{\c{c}}\else \c{c}\fi{}ois-Marie and Kirsch, Alfred and Wang, Yan and Kim, Aaram J. and Kozik, Evgeny and Stepanov, Evgeny A. and Kauch, Anna and Andergassen, Sabine and Hansmann, Philipp and Rohe, Daniel and Vilk, Yuri M. and LeBlanc, James P. F. and Zhang, Shiwei and Tremblay, A.-M. S. and Ferrero, Michel and Parcollet, Olivier and Georges, Antoine},
  journal = {Phys. Rev. X},
  volume = {11},
  issue = {1},
  pages = {011058},
  numpages = {53},
  year = {2021},
  month = {Mar},
  publisher = {American Physical Society},
  doi = {10.1103/PhysRevX.11.011058},
  url = {https://link.aps.org/doi/10.1103/PhysRevX.11.011058}
}

@article{RMP_CTQMC,
  title = {Continuous-time Monte Carlo methods for quantum impurity models},
  author = {Gull, Emanuel and Millis, Andrew J. and Lichtenstein, Alexander I. and Rubtsov, Alexey N. and Troyer, Matthias and Werner, Philipp},
  journal = {Rev. Mod. Phys.},
  volume = {83},
  issue = {2},
  pages = {349--404},
  numpages = {0},
  year = {2011},
  month = {May},
  publisher = {American Physical Society},
  doi = {10.1103/RevModPhys.83.349},
  url = {https://link.aps.org/doi/10.1103/RevModPhys.83.349}
}

@article{maier_quantum_2005-3,
  title = {Quantum Cluster Theories},
  author = {Maier, Thomas and Jarrell, Mark and Pruschke, Thomas and Hettler, Matthias H.},
  year = {2005},
  month = oct,
  journal = {Reviews of Modern Physics},
  volume = {77},
  number = {3},
  pages = {1027--1080},
  publisher = {{American Physical Society}},
  doi = {10.1103/RevModPhys.77.1027}
}

@ARTICLE{Caffarel:1994,
  author = {Caffarel, M. and Krauth, W.},
  title = {Exact diagonalization approach to correlated fermions in infinite
	dimensions: Mott transition and superconductivity},
  journal = {Phys. Rev. Lett.},
  year = {1994},
  volume = {72},
  pages = {1545},
  url = {http://link.aps.org/doi/10.1103/PhysRevLett.72.1545}
}

@article{gros_cluster_1993,
	title = {Cluster expansion for the self-energy: {A} simple many-body method for interpreting the photoemission spectra of correlated {Fermi} systems},
	volume = {48},
	shorttitle = {Cluster expansion for the self-energy},
	url = {http://link.aps.org/doi/10.1103/PhysRevB.48.418},
	doi = {10.1103/PhysRevB.48.418},
	number = {1},
	urldate = {2014-09-29},
	journal = {Physical Review B},
	author = {Gros, Claudius and Valentí, Roser},
	month = jul,
	year = {1993},
	pages = {418--425}
}

@article{senechal_spectral_2000,
	title = {Spectral {Weight} of the {Hubbard} {Model} through {Cluster} {Perturbation} {Theory}},
	volume = {84},
	url = {http://link.aps.org/doi/10.1103/PhysRevLett.84.522},
	doi = {10.1103/PhysRevLett.84.522},
	number = {3},
	urldate = {2014-09-29},
	journal = {Physical Review Letters},
	author = {Sénéchal, D. and Perez, D. and Pioro-Ladrière, M.},
	month = jan,
	year = {2000},
	pages = {522--525}
}

@article{Senechal2002,
  title = {Cluster perturbation theory for Hubbard models},
  author = {S\'en\'echal, David and Perez, Danny and Plouffe, Dany},
  journal = {Phys. Rev. B},
  volume = {66},
  issue = {7},
  pages = {075129},
  numpages = {11},
  year = {2002},
  month = {Aug},
  publisher = {American Physical Society},
  doi = {10.1103/PhysRevB.66.075129},
  url = {https://link.aps.org/doi/10.1103/PhysRevB.66.075129}
}

@article{Haule2007,
  title = {Quantum Monte Carlo impurity solver for cluster dynamical mean-field theory and electronic structure calculations with adjustable cluster base},
  author = {Haule, Kristjan},
  journal = {Phys. Rev. B},
  volume = {75},
  issue = {15},
  pages = {155113},
  numpages = {12},
  year = {2007},
  month = {Apr},
  publisher = {American Physical Society},
  doi = {10.1103/PhysRevB.75.155113},
  url = {https://link.aps.org/doi/10.1103/PhysRevB.75.155113}
}

@article{Zgid_2012,
   title={Truncated configuration interaction expansions as solvers for correlated quantum impurity models and dynamical mean-field theory},
   author={Zgid, Dominika and Gull, Emanuel and Chan, Garnet Kin-Lic},
   journal={Phys. Rev. B},
   volume={86},
   number={16},
   year={2012},
   month={Oct},
   publisher = {American Physical Society},
   url = {https://link.aps.org/doi/10.1103/PhysRevB.86.165128},   
}

@article{Go_2017,
   title={Adaptively truncated Hilbert space based impurity solver for dynamical mean-field theory},
   volume={96},
   ISSN={2469-9969},
   url={https://link.aps.org/doi/10.1103/PhysRevB.96.085139},
   number={8},
   journal={Physical Review B},
   publisher={American Physical Society (APS)},
   author={Go, Ara and Millis, Andrew J.},
   year={2017},
   month=aug }

@book{bai_templates_2000,
	address = {Philadelphia, PA},
	series = {Software, environments, tools},
	title = {Templates for the solution of algebraic eigenvalue problems},
	isbn = {978-0-89871-471-5},
	publisher = {Society for Industrial and Applied Mathematics},
	editor = {Bai, Zhaojun},
	year = {2000},
    url={https://www.netlib.org/utk/people/JackDongarra/etemplates/book.html}
}

@article{lanczos_iteration_1950,
	title = {An iteration method for the solution of the eigenvalue problem of linear differential and integral operators},
	volume = {45},
	issn = {0091-0635},
	url = {https://nvlpubs.nist.gov/nistpubs/jres/045/jresv45n4p255_A1b.pdf},
	doi = {10.6028/jres.045.026},
	number = {4},
	urldate = {2024-12-20},
	journal = {Journal of Research of the National Bureau of Standards},
	author = {Lanczos, C.},
	month = oct,
	year = {1950},
	pages = {255},
}

@article{PhysRevLett.69.2863,
  title = {Density matrix formulation for quantum renormalization groups},
  author = {White, Steven R.},
  journal = {Phys. Rev. Lett.},
  volume = {69},
  issue = {19},
  pages = {2863--2866},
  numpages = {0},
  year = {1992},
  month = {Nov},
  publisher = {American Physical Society},
  doi = {10.1103/PhysRevLett.69.2863},
  url = {https://link.aps.org/doi/10.1103/PhysRevLett.69.2863}
}

@article{VANHOUCKE201095,
title = {Diagrammatic Monte Carlo},
journal = {Physics Procedia},
volume = {6},
pages = {95-105},
year = {2010},
note = {Computer Simulations Studies in Condensed Matter Physics XXI},
issn = {1875-3892},
doi = {https://doi.org/10.1016/j.phpro.2010.09.034},
url = {https://www.sciencedirect.com/science/article/pii/S1875389210006498},
author = {Kris {Van Houcke} and Evgeny Kozik and N. Prokof’ev and B. Svistunov}
}

@article{PhysRevB.92.115109,
  title = {Mott physics and spin fluctuations: A unified framework},
  author = {Ayral, Thomas and Parcollet, Olivier},
  journal = {Phys. Rev. B},
  volume = {92},
  issue = {11},
  pages = {115109},
  numpages = {6},
  year = {2015},
  month = {Sep},
  publisher = {American Physical Society},
  doi = {10.1103/PhysRevB.92.115109},
  url = {https://link.aps.org/doi/10.1103/PhysRevB.92.115109}
}

@article{PhysRevB.75.045118,
  title = {Dynamical vertex approximation: A step beyond dynamical mean-field theory},
  author = {Toschi, A. and Katanin, A. A. and Held, K.},
  journal = {Phys. Rev. B},
  volume = {75},
  issue = {4},
  pages = {045118},
  numpages = {8},
  year = {2007},
  month = {Jan},
  publisher = {American Physical Society},
  doi = {10.1103/PhysRevB.75.045118},
  url = {https://link.aps.org/doi/10.1103/PhysRevB.75.045118}
}

@article{tubman2020,
author = {Tubman, Norm M. and Freeman, C. Daniel and Levine, Daniel S. and Hait, Diptarka and Head-Gordon, Martin and Whaley, K. Birgitta},
title = {Modern Approaches to Exact Diagonalization and Selected Configuration Interaction with the Adaptive Sampling CI Method},
journal = {Journal of Chemical Theory and Computation},
volume = {16},
number = {4},
pages = {2139-2159},
year = {2020},
doi = {10.1021/acs.jctc.8b00536},
URL = {https://doi.org/10.1021/acs.jctc.8b00536}
}

@book{sherrill1999,
title = {The Configuration Interaction Method: Advances in Highly Correlated Approaches},
editor = {Per-Olov Löwdin and John R. Sabin and Michael C. Zerner and Erkki Brändas},
series = {Advances in Quantum Chemistry},
publisher = {Academic Press},
volume = {34},
pages = {143-269},
year = {1999},
issn = {0065-3276},
doi = {https://doi.org/10.1016/S0065-3276(08)60532-8},
url = {https://www.sciencedirect.com/science/article/pii/S0065327608605328},
author = {C. {David Sherrill} and Henry F. Schaefer},
}

@article{tubman2016-2,
    author = {Tubman, Norm M. and Lee, Joonho and Takeshita, Tyler Y. and Head-Gordon, Martin and Whaley, K. Birgitta},
    title = {A deterministic alternative to the full configuration interaction quantum Monte Carlo method},
    journal = {The Journal of Chemical Physics},
    volume = {145},
    number = {4},
    pages = {044112},
    year = {2016},
    month = {07},
    issn = {0021-9606},
    doi = {10.1063/1.4955109},
    url = {https://doi.org/10.1063/1.4955109},
}

@article{mejuto2019,
  title = {Dynamical mean field theory simulations with the adaptive sampling configuration interaction method},
  author = {Mejuto-Zaera, Carlos and Tubman, Norm M. and Whaley, K. Birgitta},
  journal = {Phys. Rev. B},
  volume = {100},
  issue = {12},
  pages = {125165},
  numpages = {10},
  year = {2019},
  month = {Sep},
  publisher = {American Physical Society},
  doi = {10.1103/PhysRevB.100.125165},
  url = {https://link.aps.org/doi/10.1103/PhysRevB.100.125165}
}

@article{holmes2016,
author = {Holmes, Adam A. and Tubman, Norm M. and Umrigar, C. J.},
title = {Heat-Bath Configuration Interaction: An Efficient Selected Configuration Interaction Algorithm Inspired by Heat-Bath Sampling},
journal = {Journal of Chemical Theory and Computation},
volume = {12},
number = {8},
pages = {3674-3680},
year = {2016},
doi = {10.1021/acs.jctc.6b00407},
note ={PMID: 27428771},
URL = {https://doi.org/10.1021/acs.jctc.6b00407},
}

@article{schriber2016,
    author = {Schriber, Jeffrey B. and Evangelista, Francesco A.},
    title = {Communication: An adaptive configuration interaction approach for strongly correlated electrons with tunable accuracy},
    journal = {The Journal of Chemical Physics},
    volume = {144},
    number = {16},
    pages = {161106},
    year = {2016},
    month = {04},
    issn = {0021-9606},
    doi = {10.1063/1.4948308},
    url = {https://doi.org/10.1063/1.4948308},
}

@article{zimmerman2017,
    author = {Zimmerman, Paul M.},
    title = {Incremental full configuration interaction},
    journal = {The Journal of Chemical Physics},
    volume = {146},
    number = {10},
    pages = {104102},
    year = {2017},
    month = {03},
    issn = {0021-9606},
    doi = {10.1063/1.4977727},
    url = {https://doi.org/10.1063/1.4977727},
}

@article{zimmerman2017-2,
    author = {Zimmerman, Paul M.},
    title = {Strong correlation in incremental full configuration interaction},
    journal = {The Journal of Chemical Physics},
    volume = {146},
    number = {22},
    pages = {224104},
    year = {2017},
    month = {06},
    issn = {0021-9606},
    doi = {10.1063/1.4985566},
    url = {https://doi.org/10.1063/1.4985566},
}

@article{schriber2017,
author = {Schriber, Jeffrey B. and Evangelista, Francesco A.},
title = {Adaptive Configuration Interaction for Computing Challenging Electronic Excited States with Tunable Accuracy},
journal = {Journal of Chemical Theory and Computation},
volume = {13},
number = {11},
pages = {5354-5366},
year = {2017},
doi = {10.1021/acs.jctc.7b00725},
note ={PMID: 28892621},
URL = {https://doi.org/10.1021/acs.jctc.7b00725},
}

@article{otsuka1999,
doi = {10.1088/0954-3899/25/4/023},
url = {https://dx.doi.org/10.1088/0954-3899/25/4/023},
year = {1999},
month = {apr},
publisher = {},
volume = {25},
number = {4},
pages = {699},
author = {Takaharu Otsuka and Takahiro Mizusaki and Michio Honma},
title = {Monte Carlo shell-model calculations},
journal = {Journal of Physics G: Nuclear and Particle Physics},
}

@article{takashi2007,
  title = {Quantum Monte Carlo diagonalization for many-fermion systems},
  author = {Yanagisawa, Takashi},
  journal = {Phys. Rev. B},
  volume = {75},
  issue = {22},
  pages = {224503},
  numpages = {12},
  year = {2007},
  month = {Jun},
  publisher = {American Physical Society},
  doi = {10.1103/PhysRevB.75.224503},
  url = {https://link.aps.org/doi/10.1103/PhysRevB.75.224503}
}

@article{toyama2005,
  title = {Exact diagonalization study of optical conductivity in the two-dimensional Hubbard model},
  author = {Tohyama, T. and Inoue, Y. and Tsutsui, K. and Maekawa, S.},
  journal = {Phys. Rev. B},
  volume = {72},
  issue = {4},
  pages = {045113},
  numpages = {5},
  year = {2005},
  month = {Jul},
  publisher = {American Physical Society},
  doi = {10.1103/PhysRevB.72.045113},
  url = {https://link.aps.org/doi/10.1103/PhysRevB.72.045113}
}

@article{PhysRevB.71.241103,
  title = {Wave function optimization in the variational Monte Carlo method},
  author = {Sorella, Sandro},
  journal = {Phys. Rev. B},
  volume = {71},
  issue = {24},
  pages = {241103(R)},
  numpages = {4},
  year = {2005},
  month = {Jun},
  publisher = {American Physical Society},
  doi = {10.1103/PhysRevB.71.241103},
  url = {https://link.aps.org/doi/10.1103/PhysRevB.71.241103}
}

@article{PhysRev.138.A442,
  title = {Ground State of Liquid ${\mathrm{He}}^{4}$},
  author = {McMillan, W. L.},
  journal = {Phys. Rev.},
  volume = {138},
  issue = {2A},
  pages = {A442--A451},
  numpages = {0},
  year = {1965},
  month = {Apr},
  publisher = {American Physical Society},
  doi = {10.1103/PhysRev.138.A442},
  url = {https://link.aps.org/doi/10.1103/PhysRev.138.A442}
}

@article{PhysRevLett.51.1900,
  title = {Monte Carlo Study of the Two-Dimensional Hubbard Model},
  author = {Hirsch, J. E.},
  journal = {Phys. Rev. Lett.},
  volume = {51},
  issue = {20},
  pages = {1900--1903},
  numpages = {0},
  year = {1983},
  month = {Nov},
  publisher = {American Physical Society},
  doi = {10.1103/PhysRevLett.51.1900},
  url = {https://link.aps.org/doi/10.1103/PhysRevLett.51.1900}
}

@article{PhysRevB.31.4403,
  title = {Two-dimensional Hubbard model: Numerical simulation study},
  author = {Hirsch, J. E.},
  journal = {Phys. Rev. B},
  volume = {31},
  issue = {7},
  pages = {4403--4419},
  numpages = {0},
  year = {1985},
  month = {Apr},
  publisher = {American Physical Society},
  doi = {10.1103/PhysRevB.31.4403},
  url = {https://link.aps.org/doi/10.1103/PhysRevB.31.4403}
}

@article{10.1021/acs.jctc.7b00682,
    author = {Ronca, Enrico and Li, Zhendong and Jimenez-Hoyos, Carlos A. and Chan, Garnet Kin-Lic},
    title = {Time-Step Targeting Time-Dependent and Dynamical Density
Matrix Renormalization Group Algorithms with ab Initio Hamiltonians},
    journal = {Journal of Chemical Theory and Computation},
    volume = {13},
    number = {11},
    pages = {5560-5571},
    year = {2017},
    month = {09},
    issn = {1549-9618},
    doi = {10.1021/acs.jctc.7b00682},
    url = {https://doi.org/10.1021/acs.jctc.7b00682},
    eprint = {https://pubs.acs.org/jctcce/article-pdf/13/11/5560/6919449/ct7b00682.pdf}
}

@misc{pyblock3,
    author = {Huanchen, Zhai and Yang, Gao and Garnet, K.-L. Chan.},
    title = {\texttt{pyblock3}: an efficient python block-sparse tensor and MPS/DMRG library.},
    year = {2021},
    link = {github.com/block-hczhai/pyblock3-preview},
    url = {https://github.com/block-hczhai/pyblock3-preview}
}
\end{document}